\documentclass[aps,prl,showpacs,twocolumn,preprintnumbers,superscriptaddress,nofootinbib,floatfix,10pt]{revtex4-1}
\usepackage{amsfonts,amssymb,stmaryrd,latexsym,amsmath}
\usepackage{subfigure,graphicx}
\usepackage{times}
\usepackage{slashed}

\begin{document}

\title{\textbf{\textit{Ab initio}} calculations of the isotopic dependence of nuclear clustering }

\author{Serdar~Elhatisari}
\affiliation{Helmholtz-Institut f\"ur Strahlen- und
             Kernphysik and Bethe Center for Theoretical Physics,
             Universit\"at Bonn,  D-53115 Bonn, Germany}
\affiliation{Department of Physics, Karamanoglu Mehmetbey University,
Karaman 70100, Turkey}

\author{Evgeny~Epelbaum}
\affiliation{Institut~f\"{u}r~Theoretische~Physik~II,~Ruhr-Universit\"{a}t~Bochum,
D-44870~Bochum,~Germany}
\affiliation{Kavli Institute for Theoretical Physics, University of California,
Santa Barbara, CA 93106-4030, USA}

\author{Hermann~Krebs}
\affiliation{Institut~f\"{u}r~Theoretische~Physik~II,~Ruhr-Universit\"{a}t~Bochum,
D-44870~Bochum,~Germany}
\affiliation{Kavli Institute for Theoretical Physics, University of California,
Santa Barbara, CA 93106-4030, USA}

\author{Timo~A.~L\"{a}hde}
\affiliation{Institute~for~Advanced~Simulation, Institut~f\"{u}r~Kernphysik,
and
J\"{u}lich~Center~for~Hadron~Physics,~Forschungszentrum~J\"{u}lich,
D-52425~J\"{u}lich, Germany}

\author{Dean~Lee}
\affiliation{National Superconducting Cyclotron Laboratory, Michigan State University, MI 48824, USA}
\affiliation{Department~of~Physics, North~Carolina~State~University, Raleigh,
NC~27695, USA}
\affiliation{Kavli Institute for Theoretical Physics, University of California,
Santa Barbara, CA 93106-4030, USA}
             
\author{Ning~Li$^*$} 
\affiliation{Institute~for~Advanced~Simulation, Institut~f\"{u}r~Kernphysik,
and
J\"{u}lich~Center~for~Hadron~Physics,~Forschungszentrum~J\"{u}lich,
D-52425~J\"{u}lich, Germany}

\author{Bing-nan~Lu$^\dagger$}
\affiliation{Institute~for~Advanced~Simulation, Institut~f\"{u}r~Kernphysik,
and
J\"{u}lich~Center~for~Hadron~Physics,~Forschungszentrum~J\"{u}lich,
D-52425~J\"{u}lich, Germany}

\author{Ulf-G.~Mei{\ss}ner}  
\affiliation{Helmholtz-Institut f\"ur Strahlen- und
             Kernphysik and Bethe Center for Theoretical Physics,
             Universit\"at Bonn,  D-53115 Bonn, Germany}  
\affiliation{Institute~for~Advanced~Simulation, Institut~f\"{u}r~Kernphysik,
and
J\"{u}lich~Center~for~Hadron~Physics,~Forschungszentrum~J\"{u}lich,
D-52425~J\"{u}lich, Germany}     
\affiliation{JARA~-~High~Performance~Computing, Forschungszentrum~J\"{u}lich,
D-52425 J\"{u}lich,~Germany}

\author{Gautam~Rupak}
\affiliation{Department~of~Physics~and~Astronomy and HPC$^2$ Center for Computational
Sciences, Mississippi~State~University,
Mississippi State, MS~39762, USA}

\begin{abstract}  
 Nuclear clustering describes the appearance of structures resembling smaller nuclei such as alpha particles ($^4$He nuclei) within the interior of a larger nucleus. While clustering is important for several well-known examples 
\cite{Horiuchi:2012a,Beck:2014a,Funaki:2015uya,Freer:2017gip}, much remains to be discovered about the general nature of clustering in nuclei.  In this letter we present lattice Monte Carlo calculations based on chiral effective field theory for the ground states of helium, beryllium, carbon, and oxygen isotopes.  By computing model-independent measures that probe three- and four-nucleon correlations at short distances, we determine the shape of the alpha clusters and the entanglement of nucleons comprising each alpha cluster with the outside medium.  We also introduce a new computational approach called the pinhole algorithm, which solves a long-standing deficiency of auxiliary-field Monte Carlo simulations in computing density correlations relative to the center of mass. We use the pinhole algorithm to determine the proton and neutron density distributions and the geometry of cluster correlations in $^{12}$C, $^{14}$C, and $^{16}$C.  The structural similarities among the carbon isotopes suggest that $^{14}$C and $^{16}$C have excitations analogous to the well-known Hoyle state resonance in $^{12}$C \cite{Hoyle:1954zz,Cook:1957}.           
\end{abstract}

\maketitle

There have been many exciting recent advances in {\it ab initio} nuclear
structure theory \cite{Romero-Redondo:2014fya,Maris:2014hga,Dytrych:2016vjy,Duguet:2016wwr,
Stroberg:2016ung,Ruiz:2016gne,Hagen:2016uwj,Elhatisari:2016owd} which link
nuclear forces to nuclear structure in impressive agreement with experimental data.  However, we still know very little about the quantum correlations among nucleons that give rise to nuclear clustering and collective behavior.
The main difficulty in studying alpha clusters in nuclei is that the calculation must include four-nucleon correlations.  Unfortunately in many cases this dramatically increases the amount of computer memory and computing time needed in calculations of heavier nuclei.  Nevertheless there is promising work in progress using the symmetry-adapted no-core shell model \cite{Launey:2016fvy}, antisymmetrized molecular dynamics \cite{Yoshida:2016dzr}, fermionic molecular dynamics \cite{Feldmeier:2016zut}, the alpha-container model \cite{Schuck:2017jtw}, Monte Carlo shell model \cite{Yoshida:2014tta}, and Green's function Monte Carlo \cite{Lovato:2016gkq}.

\begin{figure}[!t]
\centering
\caption {In panel {\bf a} we show the ground state energies versus number of nucleons $A$ for the hydrogen, helium, beryllium, carbon, and oxygen isotopes. The errors are one-standard deviation error bars associated with the stochastic errors and the extrapolation to an infinite number of time steps.  In panel {\bf b} we show $\rho_3/\rho_{3,\alpha}$
and $\rho_4/\rho_{4,\alpha}$ for the neutron-rich helium, beryllium, carbon, oxygen isotopes.
The error bars denote one standard deviation errors associated with
the stochastic errors and the extrapolation to an infinite number of time
steps. For comparison we show also the number of alpha clusters, $N_\alpha$. \bigskip}
\centering
\begin{subfigure}
\centering
\includegraphics[scale=0.33]{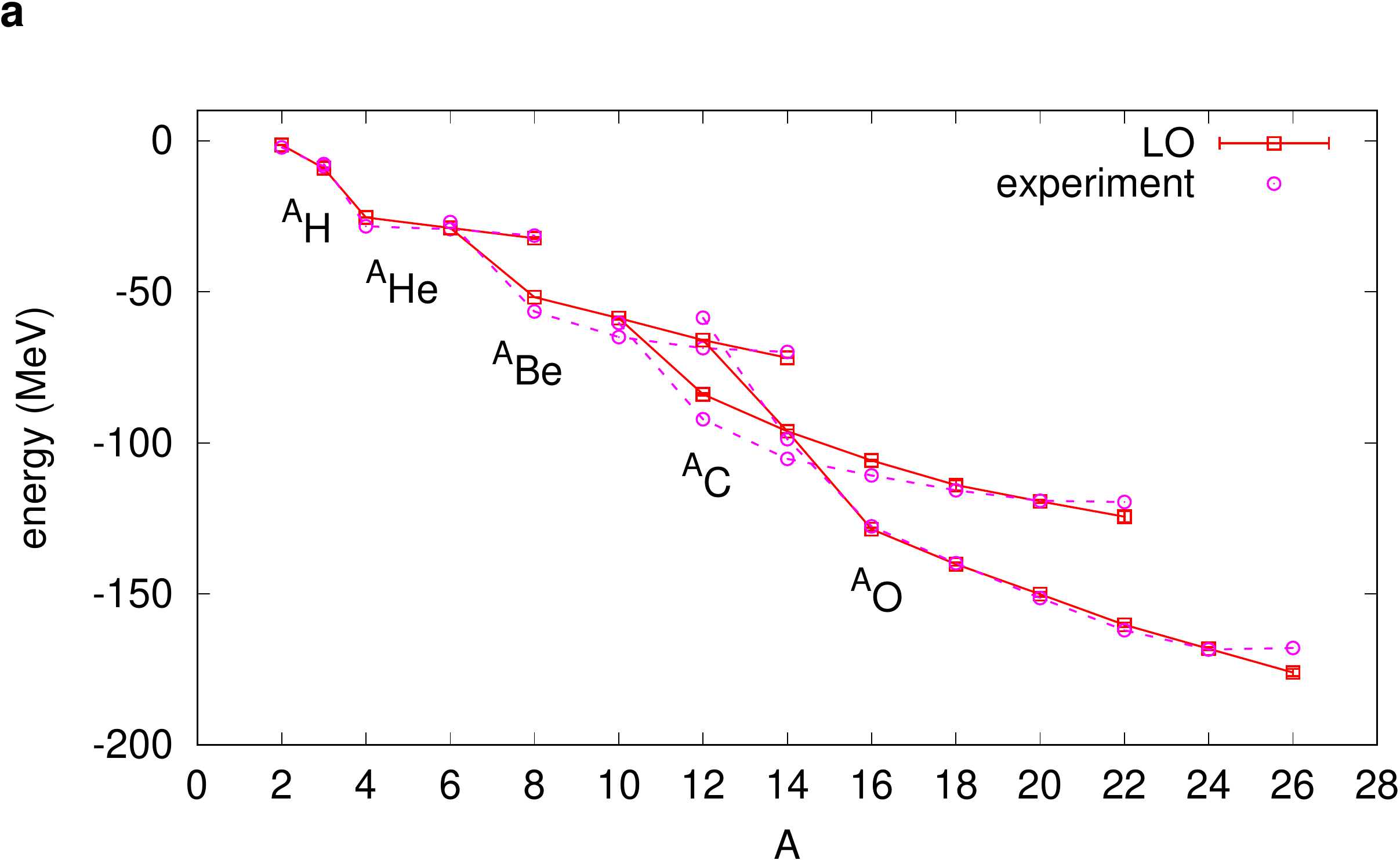}
\end{subfigure}
\begin{subfigure}
\centering
\includegraphics[scale=0.33]{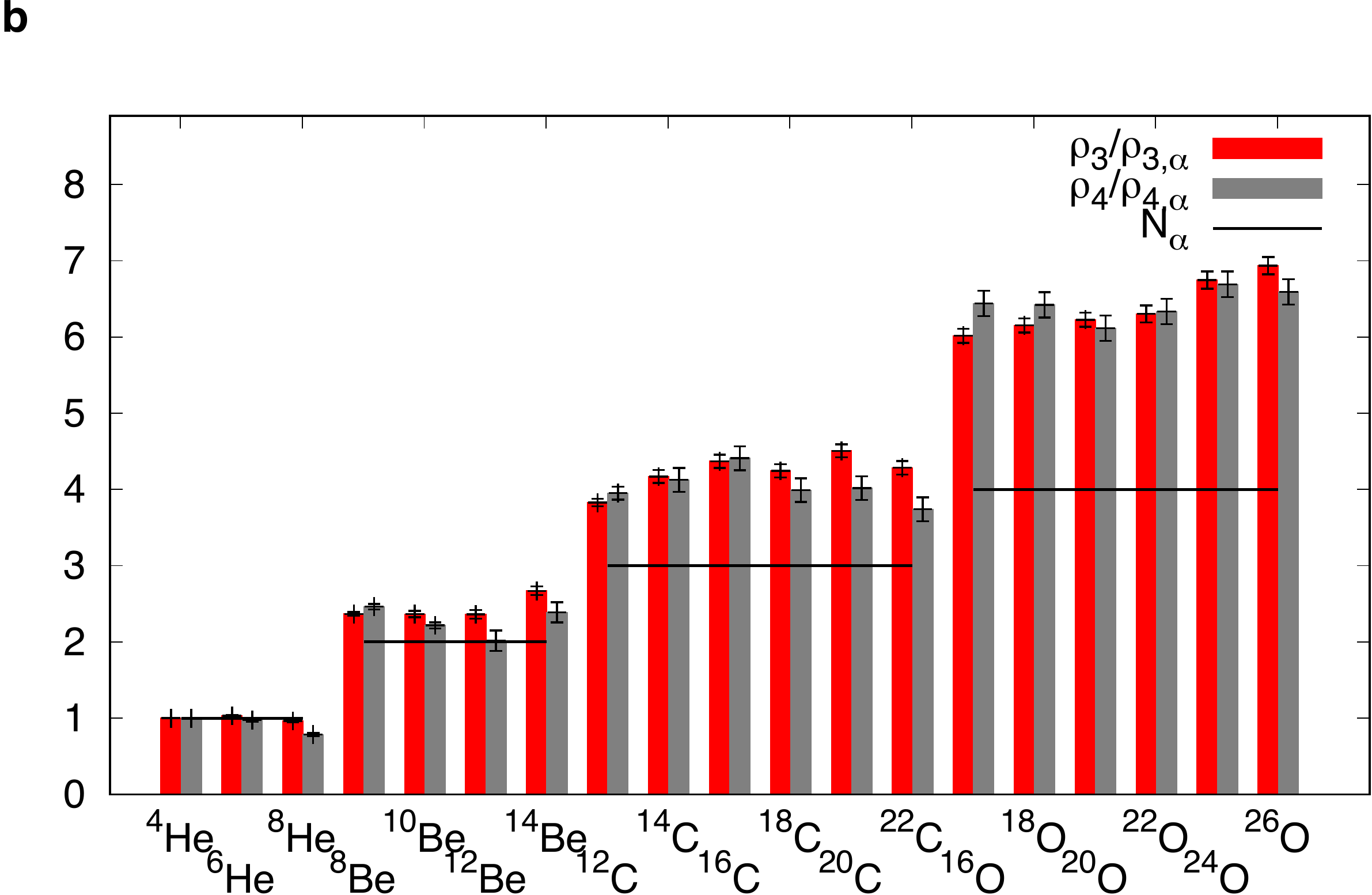}
\end{subfigure}
\label{LO_energies_new}
\end{figure}

Lattice calculations using chiral effective
field theory and auxiliary-field Monte Carlo methods have probed alpha clustering in the $^{12}$C and $^{16}$O systems \cite{Epelbaum:2011md,Epelbaum:2012qn,Epelbaum:2012iu,Epelbaum:2013paa}.  However these lattice simulations have encountered severe Monte Carlo sign oscillations in cases where the number of protons $Z$ and number of neutrons $N$ are different.  In this work we solve this problem by using a new leading-order lattice action that retains a greater amount of symmetry, thereby removing nearly all of the Monte Carlo sign oscillations. The relevant symmetry is Wigner's SU(4) spin-isospin symmetry \cite{Wigner:1937}, where the four nucleon degrees of freedom can be rotated as four components of a complex vector.  Previous attempts using SU(4) symmetry had failed due to the tendency of nuclei to overbind in larger nuclei. However recent progress has uncovered   important connections between local interactions and nuclear binding, as well as the significance of the alpha-alpha interaction 
\cite{Elhatisari:2015iga,Elhatisari:2016owd,Rokash:2016tqh}.  Following this approach, we have constructed a leading-order lattice action with highly-suppressed sign oscillations and which reproduces the ground-state binding energies of the hydrogen, helium, beryllium, carbon, and oxygen isotopes to an accuracy of 0.7 MeV per nucleon or better. The lattice results are
shown in panel {\bf a} of Fig.~\ref{LO_energies_new} in comparison with the observed ground
state energies.  The astonishingly good agreement at leading order in chiral effective field theory with only three free parameters is quite remarkable and bodes well for future calculations at higher orders. We use auxiliary-field Monte
Carlo simulations with a spatial lattice spacing of 1.97~fm and lattice time
spacing 1.97~fm$/c$. We comment that the results for these ground state energies are equally good when including Coulomb repulsion and a slightly more attractive nucleon-nucleon short-range interaction.  The full details of the lattice interaction,  nucleon-nucleon phase shifts, simulation methods, and results are given in the Supplemental Materials. 
 
Let $\rho(\bf n)$ be the total nucleon density operator on lattice site $\bf n$. We will use short-distance three- and four-nucleon operators as probes of the nuclear clusters. To construct a probe for alpha clusters, we define $\rho_4$ as the expectation value of $:\rho^4({\bf n})/4!:$
summed over $\bf n$.
The $::$ symbols indicate normal-ordering where all annihilation operators
are moved to the right and all creation operators are moved to the left.
For nuclei with even $Z$ and even $N$, there
are likely no well-defined $^3$H or $^3$He clusters since their formation is not energetically
favorable.  Therefore we can use short-distance three-nucleon operators as a second probe of alpha clusters. We define $\rho_3$
as the expectation value of $:\rho^3({\bf n})/3!:$
summed over $\bf n$. A $^3$H or $^3$He cluster may form in nuclei with odd $Z$ or odd $N$.  In these cases we can use spin- and isospin-dependent three-nucleon operators to probe the $^3$H and $^3$He clusters.   As we consider only nuclei with even $Z$ and even $N$ here, we focus on $\rho_3$ and $\rho_4$ for the remainder of the discussion.  We note that another measure of clustering in nuclei by measuring short-distance correlations has been introduced in Ref.~\cite{Zhang:2016vwy}.

Due to divergences at short distances, $\rho_3$ and $\rho_4$ will depend on the short-distance regularization scale, which in our case is the lattice spacing.  However the regularization-scale dependence of $\rho_3$ and $\rho_4$ does not depend on the nucleus being considered.  Therefore if we let $\rho_{3,\alpha}$ and $\rho_{4,\alpha}$ be the corresponding values for the alpha particle, then the ratios $\rho_3/\rho_{3,\alpha}$ and $\rho_4/\rho_{4,\alpha}$ are free from short-distance divergences and are model-independent quantities up to contributions from higher-dimensional operators in an operator product expansion. The derivations of these statements are given in the supplemental materials. We have computed $\rho_3$ and $\rho_4$ for the helium, beryllium, carbon, and oxygen isotopes.  As our leading-order
interactions are invariant under an isospin mirror flip that interchanges
protons and neutrons, we focus here on neutron-rich nuclei. The results for $\rho_3/\rho_{3,\alpha}$
and $\rho_4/\rho_{4,\alpha}$ are presented in panel {\bf b} of 
Fig.~\ref{LO_energies_new}. As we might expect, the values for $\rho_3/\rho_{3,\alpha}$
and $\rho_4/\rho_{4,\alpha}$ are roughly the
same for the different neutron-rich isotopes of each element.         

Since $\rho_4$ involves four nucleons, it couples to the center
of the alpha cluster while $\rho_3$ gets a contribution
from a wider portion of the alpha-cluster wave function.  Therefore, a value larger than
1 for the ratio  of $\rho_4/\rho_{4,\alpha}$ to $\rho_3/\rho_{3,\alpha}$  corresponds to a more compact alpha-cluster shape than in vacuum, and a value less than 1 corresponds to a more diffuse alpha-cluster shape.  In panel {\bf b} of 
Fig.~\ref{LO_energies_new} we observe that the ratio of $\rho_4/\rho_{4,\alpha}$ to $\rho_3/\rho_{3,\alpha}$ starts at 1 or slightly above  1 when $N$ is comparable to $Z$, and the ratio gradually decreases as the number of neutrons is increased. This is evidence for the swelling of the alpha clusters as the system becomes saturated with excess neutrons. The effect has also been seen in $^6$He and $^8$He in Green's Function Monte Carlo calculations \cite{Wiringa:2000gb}. 

We comment here that if one wants to study the swelling of alpha clusters in detail, then there are other local operators that provide more direct geometrical information such as the operators $:\!\rho^3({\bf n})\rho({\bf n}')\!:$ and $:\!\rho^2({\bf n})\rho^2({\bf
n}')\!:$, where ${\bf n}'$ is a nearest-neighbor site to ${\bf n}$.  These local operators have the advantage of measuring four-nucleon correlations directly rather than inferring them from the ratio of four-body and three-body correlations, which may not work well for cases with very large isospin imbalance. 

The traditional approach to nuclear clustering usually involves a variational
ansatz where the wave function is expanded in terms of  some chosen set of
alpha-cluster wave functions.  However the answer obtained this way may depend
strongly on the details of the interactions and the choice of alpha-cluster
wave functions.  This problem
of model dependence is solved by calculating short-range multi-nucleon
quantities.  Even though we use only short-range operators, the quantities
$\rho_3/\rho_{3,\alpha}$ and $\rho_4/\rho_{4,\alpha}$
act as high-fidelity alpha-cluster detectors.  Their values are strongly
enhanced if the nuclear 
wave function has a well-defined alpha-cluster substructure.  As shown in
the supplemental materials, the enhancement factor for $\rho_3/\rho_{3,\alpha}$
is $(R_A/R_\alpha)^6$, where $R_A$ is the nuclear radius and $R_\alpha$ is
the alpha-particle radius.  The enhancement factor for $\rho_4/\rho_{4,\alpha}$
is an even larger factor of $(R_A/R_\alpha)^9$. 

We denote the number of alpha clusters as $N_{\alpha}$.  A simple counting of protons gives $N_{\alpha}=1$ for neutron-rich helium, $N_{\alpha}=2$ for neutron-rich beryllium, $N_{\alpha}=3$ for neutron-rich carbon, and $N_{\alpha}=4$ for neutron-rich oxygen.  However the alpha clusters are immersed in a complex many-body system, and it is useful to quantify the entanglement of the nucleons comprising each alpha cluster with the outside medium.  The observables $\rho_3/\rho_{3,\alpha}$ and $\rho_4/\rho_{4,\alpha}$ are useful for this purpose. Let us define $\delta^{\rho_3}_{\alpha}$ as the difference $\rho_3/\rho_{3,\alpha}-N_{\alpha}$ divided by $N_{\alpha}$. Since $\delta^{\rho_3}_\alpha$ measures the deviation of the nuclear
wave function from a pure product state of alpha clusters and excess nucleons, we call it the $\rho_3$-entanglement of the alpha clusters.
 In an analogous manner, we can also define the $\rho_4$-entanglement 
 $\delta^{\rho_4}_{\alpha}$ as the difference $\rho_4/\rho_{4,\alpha}-N_{\alpha}$ divided by $N_{\alpha}$. $\delta^{\rho_4}_{\alpha}$ turns out to be quantitatively similar to $\delta^{\rho_3}_{\alpha}$, though with more sensitivity to the shape of the alpha clusters.

In panel {\bf b} of 
Fig.~\ref{LO_energies_new}, we show $N_{\alpha}$ along with the ratios 
$\rho_3/\rho_{3,\alpha}$ and $\rho_4/\rho_{4,\alpha}$.  The relative excess of $\rho_3/\rho_{3,\alpha}$ compared to $N_{\alpha}$ gives $\delta^{\rho_3}_\alpha$, and the relative excess of $\rho_4/\rho_{4,\alpha}$ compared to $N_{\alpha}$ gives $\delta^{\rho_4}_\alpha$.  We see that $\delta^{\rho_3}_\alpha$ is negligible for $^6$He and $^8$He, indicating an almost pure product state of alpha clusters and excess neutrons.  For the beryllium isotopes, $\delta^{\rho_3}_\alpha$ is about $0.18$\footnote{In this leading-order calculation the $^8$Be ground state is about 1 MeV below the two-$\alpha$ threshold.  The addition of the Coulomb interaction and other corrections should push this energy closer to threshold, and one expects $\delta^{\rho_3}_\alpha$ to decrease as a result.} for $^8$Be and rises to about $0.34$ for $^{14}$Be.  For the carbon isotopes, it is about $0.28$ for $^{12}$C and rises to a maximum of about $0.50$ near the drip line. For the
oxygen isotopes, $\delta^{\rho_3}_\alpha$ is about $0.50$ for $^{16}$O and increases with neutron number up to $0.73$.  For such high values of the $\rho_3$-entanglement, we expect a simple picture in terms of alpha clusters and excess neutrons will break down. $\delta^{\rho_3}_\alpha$
should be much lower for excited cluster-like states of the oxygen isotopes.
With $\rho_3$-entanglement, we have a model-independent quantitative
measure of nuclear cluster formation
in terms of entanglement of the wave function.
Our results show that
the transition from cluster-like states in light systems to nuclear
liquid-like states in heavier systems should not be viewed as a simple suppression
of multi-nucleon short-distance correlations, but rather an increasing
entanglement of the nucleons involved in the multi-nucleon
correlations. 
 
Despite the many computational advantages of auxiliary-field Monte Carlo methods, one fundamental deficiency is that the simulations involve quantum states that are superpositions of many different center-of-mass positions.  Therefore density distributions of the  
nucleons cannot be computed directly.  To solve this problem we have developed a new method called the pinhole algorithm.  In this algorithm an opaque screen is placed at the middle time step with pinholes bearing spin and isospin  labels that allow nucleons with the corresponding spin and isospin to pass.  We use $A$ pinholes for a simulation of $A$ nucleons,  and the locations as well as the spin and isospin labels of the pinholes are updated by Monte Carlo importance sampling.  From the simulations, we obtain the expectation value of the normal-ordered $A$-body density operator $:\rho_{i_1,j_1}({\bf n}_1)\cdots \rho_{i_A,j_A}({\bf
n}_A):$, where $\rho_{i,j}$ is the density operator for a nucleon with spin $i$ and isospin $j$. 

\begin{figure}[!ht]
\centering
\caption {Plots of the proton and neutron densities for the ground states of $^{12}$C, $^{14}$C, and $^{16}$C versus radial distance.   We show data for $L_t=7,9,11,13,15$ time steps. We show $^{12}$C in panel {\bf a}, $^{14}$C in panel {\bf b}, and $^{16}$C in panel {\bf c}. The errors are one-standard deviation error bars associated with the stochastic
errors. For comparison we show the experimentally observed proton densities for $^{12}$C and $^{14}$C
\cite{Kline:1973pi}. \bigskip}
\centering
\includegraphics[scale=0.45]{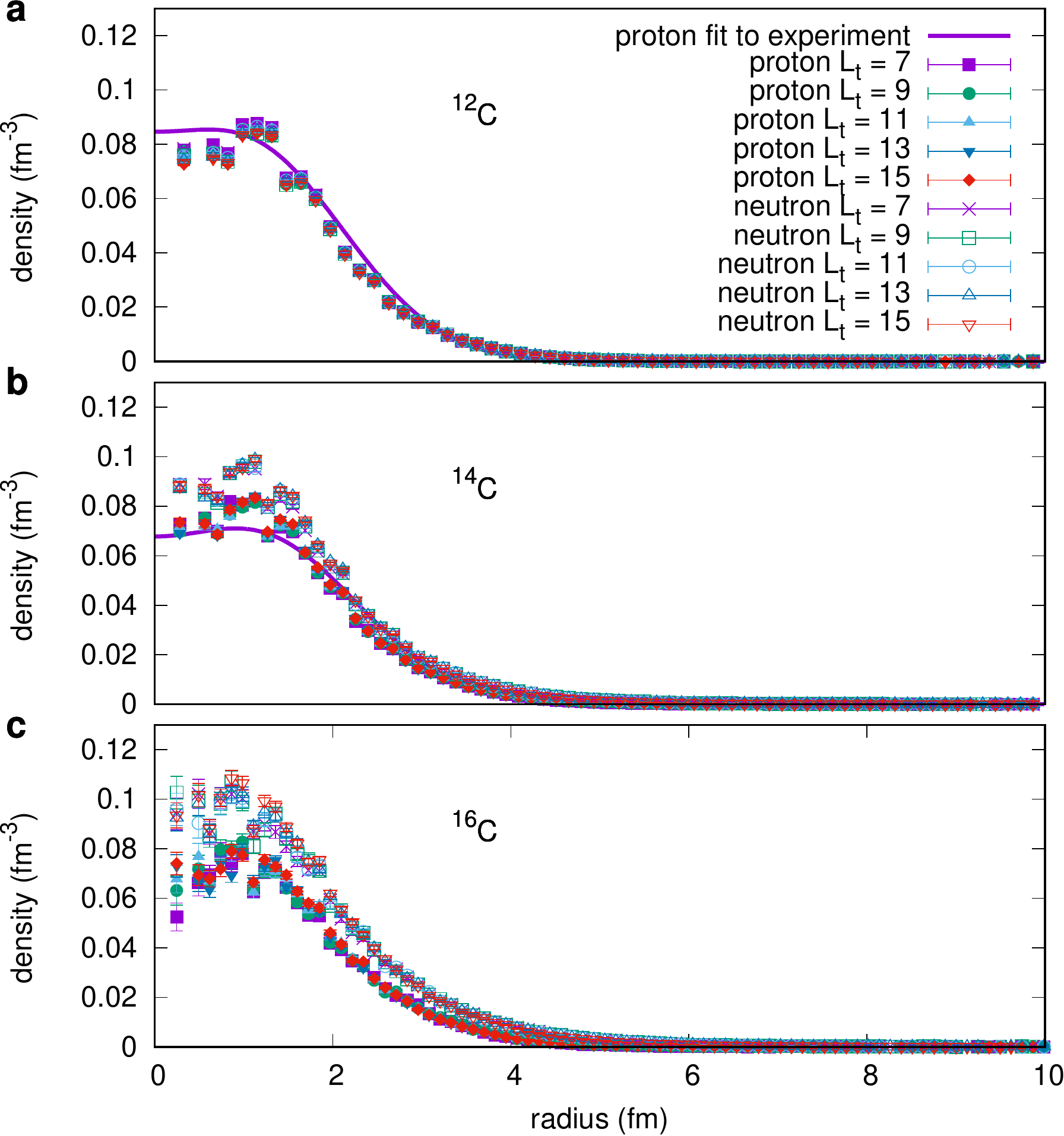}
\label{density}
\end{figure}

Using the pinhole algorithm, we have computed the proton and neutron densities for the ground states of $^{12}$C, $^{14}$C, and $^{16}$C.  In order to account for the nonzero size of the nucleons, we have convolved the point-nucleon distributions with a Gaussian distribution with root-mean-square radius 0.84 fm, the charge radius of the proton \cite{Belushkin:2006qa,Pohl:2010zza}.  The results are shown in Fig.~\ref{density} along with the experimentally observed proton densities for $^{12}$C and $^{14}$C \cite{Kline:1973pi}, which we define as the charge density divided by the electric charge $e$.
From Fig.~\ref{density} we see that the agreement between the calculated proton densities and experimental data for $^{12}$C and $^{14}$C is rather good. We show data for $L_t=7,9,11,13,15$ time steps.  The fact that the results have little dependence on $L_t$ means that we are seeing ground state properties. As we increase the number of neutrons and go from $^{12}$C to $^{16}$C, the shape of the proton density profile remains roughly the same.  However there is a gradual decrease in the central density and a broadening of the proton density distribution.  We see also that the excess neutrons in $^{14}$C and $^{16}$C are distributed fairly evenly, appearing in both the central region as well as the tail.

We now study the alpha-cluster structures of $^{12}$C, $^{14}$C, and $^{16}$C in more detail.  In order to probe the alpha cluster geometry, we use the fact that there is only one spin-up proton per alpha cluster.  Using the pinhole algorithm, we consider the triangular shapes formed by the three spin-up protons in the carbon isotopes.  This correlation function is free of short-distance divergences, and so, up to the contribution of higher-dimensional operators, it provides a model-independent measure that serves as a proxy for the geometry of the alpha-cluster configurations. 

\begin{figure}[!ht]
\caption {The two red spheres with arrows indicate the first two spin-up protons, and
the line connecting them is the longest side of the triangle.  We show the third spin-up proton probability distribution in $^{12}$C in panel {\bf a}, $^{14}$C in panel {\bf b}, and $^{16}$C in panel {\bf c}. The results are computed
at $L_t=15$ time steps. In  panel {\bf d} we show the
third
spin-up proton probability distribution for a simple Gaussian lattice model
of the distribution of the spin-up protons.}

\label{triangle1}
\centering
\begin{subfigure}
\centering
\includegraphics[scale=0.21,trim={0 0 0 0cm},clip]{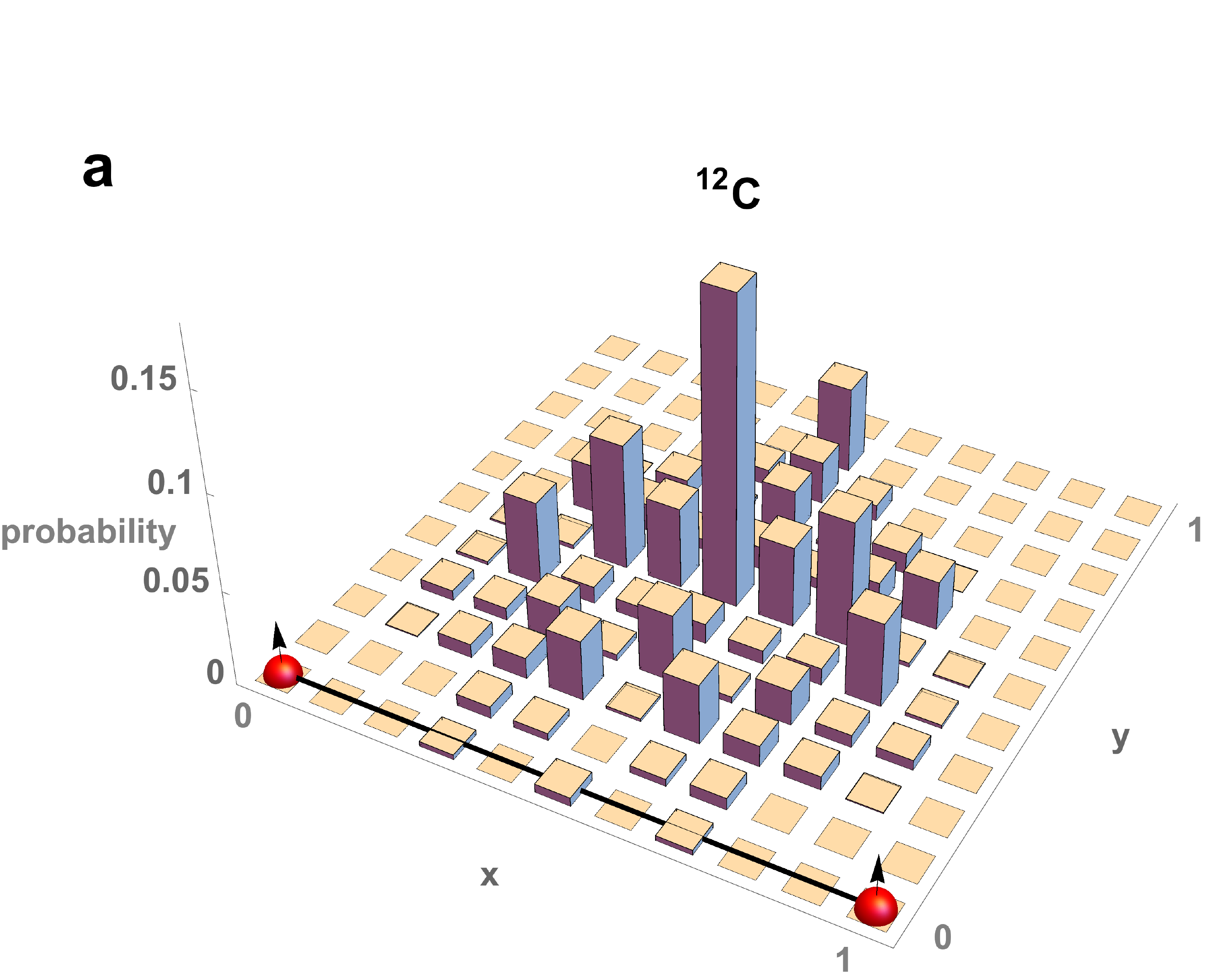}
\end{subfigure}
\begin{subfigure}
\centering
\includegraphics[scale=0.21,trim={0 0 0 3cm},clip]{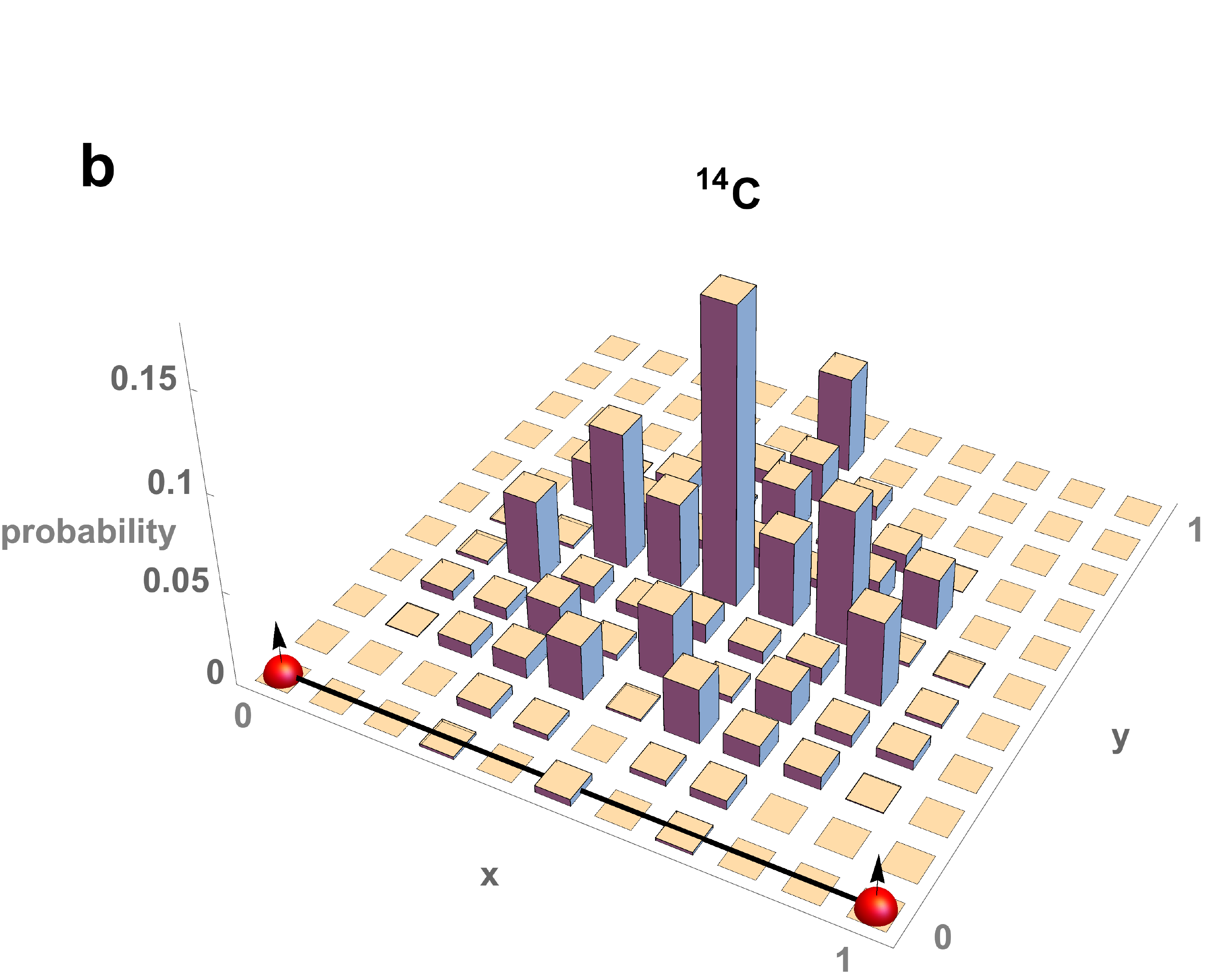}
\end{subfigure}

\centering
\begin{subfigure}
\centering
\includegraphics[scale=0.21,trim={0 0 0 3cm},clip]{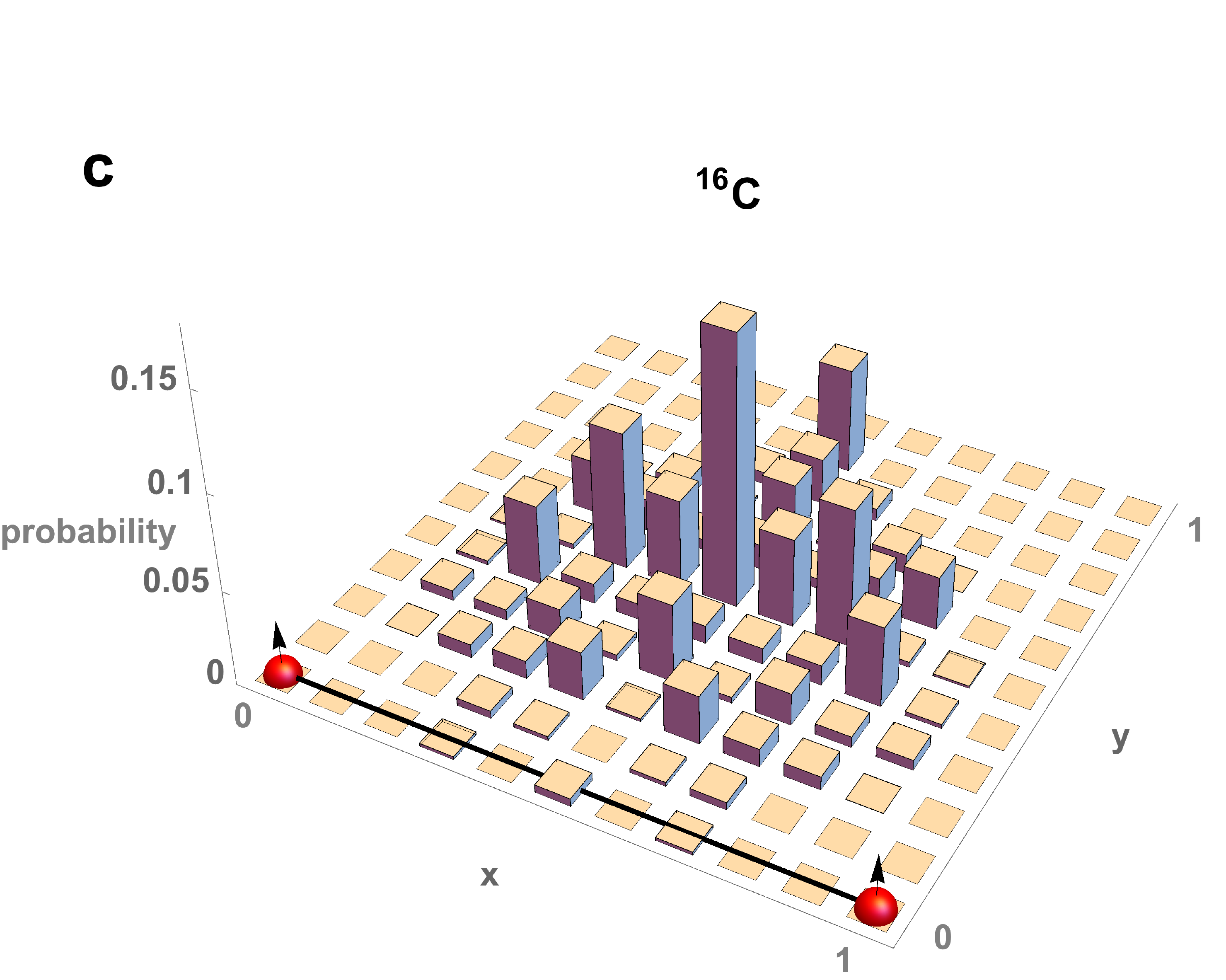}
\end{subfigure}
\begin{subfigure}
\centering
\includegraphics[scale=0.21,trim={0 0 0 3cm},clip]{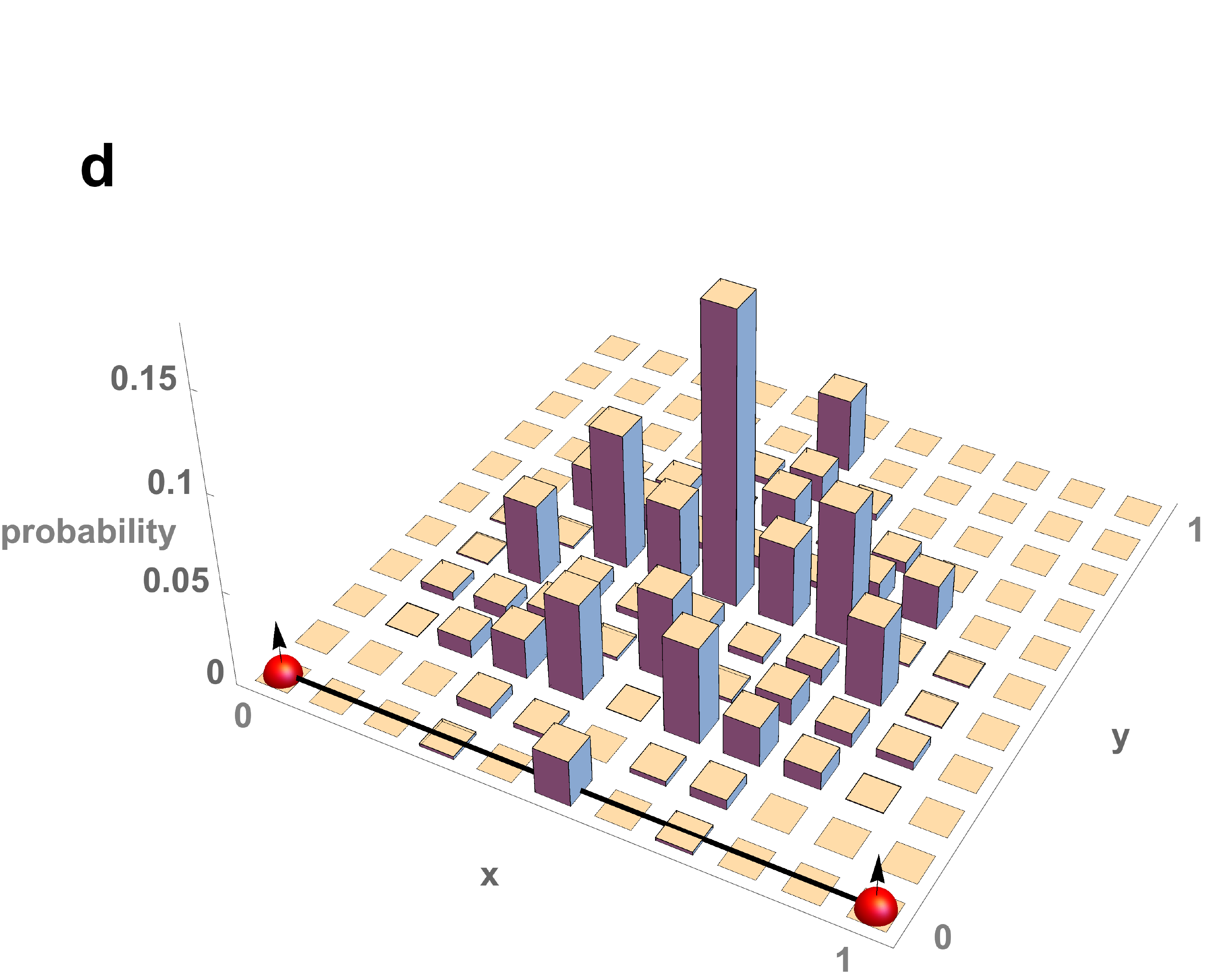}
\end{subfigure}
\end{figure}

The three spin-up protons form the vertices of a triangle.  When collecting the lattice simulation data, we rotate the triangle so that the longest side lies on the $x$-axis.  We also rescale the triangle so the longest side has length one, and flip the triangle, if needed, so that the third spin-up
proton is in the upper half of the $xy$-plane.  Histograms of the third spin-up proton probability distributions for   $^{12}$C, $^{14}$C, and $^{16}$C are plotted in panel {\bf a}, {\bf b}, {\bf c} of Fig.~\ref{triangle1} using the data at $L_t=15$ time steps. 
The data for other values of $L_t$ are almost identical.  There is some jaggedness due to the discreteness of the lattice, but we see quite clearly that the  histograms for  $^{12}$C, $^{14}$C, and $^{16}$C are very similar.  While there is some increase in the overall radius of the nucleus, the rescaled cluster geometry of the three carbon isotopes remain largely the same.  In each case we see that there is a strong preference for
triangles where the largest angle is less than or equal to 90 degrees.
We should note that idea that the ground state of $^{12}$C has an acute triangular alpha-cluster structure has a long history dating back to Ref.~\cite{Hafstad:1938}.

Given the rich cluster structure of the excited states of $^{12}$C, this raises the interesting possibility of similar cluster states appearing in $^{14}$C and $^{16}$C.  In particular, the bound $0^+_2$ state at 6.59~MeV above the ground state of $^{14}$C may be a bound-state analog to the Hoyle state resonance in $^{12}$C at 7.65~MeV.  It may also have a clean experimental signature since low-lying neutron excitations are suppressed by the shell closure at eight neutrons.  There is also a bound $0^+_2$ in $^{16}$C, however in this case one expects low-lying two-neutron excitations to be important, thereby making the analysis more complicated. We note that there is ample experimental evidence for the cluster properties
of the neutron-rich beryllium and carbon isotopes 
\cite{Bohlen:2007ket,Bohlen:2003ktf,Freer:2008pea,Marin-Lambarri:2014zxa}.

In order to analyze what we are seeing in the lattice data, we can make 
a simple Gaussian lattice model of the distribution of the spin-up protons.
 We consider a probability distribution
$P({\bf {r}_1},{\bf {r}_2},{\bf {r}_3})$ on our lattice grid for the positions
of the protons ${\bf {r}_1}$, ${\bf {r}_2}$, and ${\bf {r}_3}$.  We take
the probability
distribution to be a product of Gaussians with root-mean-square radius 2.6~fm
(charge radius of $^{14}$C) and unit step functions which vanish if the magnitude
of ${\bf {r}_1}-{\bf {r}_2}$, ${\bf {r}_2}-{\bf {r}_3}$, or ${\bf {r}_3}-{\bf
{r}_1}$ is smaller than 1.7 fm (charge radius of $^4$He),  

\begin{equation}
 \exp\left[-\frac{\sum_{i}{\bf {r}_i}^2}{2(2.6\,\rm{fm})^2} \right]\prod_{j> k}\theta(|{\bf {r}_j-\bf {r}_k}|-1.7\,\rm{fm}).
\end{equation}We can factor
out the
center-of-mass distribution of the three spin-up protons and recast the Gaussian
factors as a product of Gaussians for the separation vectors ${\bf {r}_1}-{\bf
{r}_2}$, ${\bf {r}_2}-{\bf {r}_3}$, or ${\bf {r}_3}-{\bf
{r}_1}$ with root-mean-square radius
4.5~fm,  
\begin{equation}
\prod_{j> k}\exp\left[-\frac{{(\bf {r}_j-\bf {r}_k)}^2}{2(4.5\,\rm{fm})^2}
\right]\theta(|{\bf {r}_j-\bf {r}_k}|-1.7\,\rm{fm}).
\end{equation}
In panel {\bf d} of Fig.~\ref{triangle1} we show the third
spin-up proton probability distribution corresponding to this model. Despite
the simplicity of this model with no free parameters, we note the good agreement
with the lattice data for $^{12}$C, $^{14}$C, and $^{16}$C.  The only discrepancy
is that the model overpredicts the probability of producing obtuse triangular
configurations. This indicates that there are some additional correlations
between the clusters that go beyond this simple Gaussian lattice model.

In this letter we have presented a number of novel approaches to computing and quantifying clustering and entanglement in nuclei.  We hope that this work may help to accelerate progress in theoretical and experimental efforts to understand the correlations that produce nuclear clustering and collective behavior. 

\section*{Acknowledgement}
We are grateful for the hospitality of the Kavli Institute for Theoretical Physics
at UC Santa Barbara for hosting E.E., H.K., and D.L.  We are indebted to Ingo Sick for providing the experimental data tables on the electric form factor for $^{12}$C. We acknowledge partial financial support
from the CRC110: Deutsche Forschungsgemeinschaft (SGB/TR 110, ``Symmetries and the Emergence of Structure in QCD''), the BMBF (Verbundprojekt 05P2015 - NUSTAR R\&D),
the U.S. Department of Energy (DE-FG02-03ER41260),
and U.S. National Science Foundation grant No. PHY-1307453.
Further support
was provided by the Magnus Ehrnrooth Foundation of the Finnish Society of Sciences
and Letters and the Chinese Academy of Sciences (CAS) President's International Fellowship Initiative (PIFI) grant no. 2017VMA0025.
The computational resources 
were provided by the J\"{u}lich Supercomputing Centre at Forschungszentrum
J\"{u}lich, RWTH Aachen, and North Carolina State University.

\renewcommand{\thefigure}{S\arabic{figure}}
\setcounter{figure}{0}

\renewcommand{\thefigure}{S\arabic{figure}}
\setcounter{figure}{0}

\onecolumngrid

\newpage
\section*{Supplemental Materials}

\subsection*{Lattice interactions}
In our lattice simulations the spatial lattice spacing is taken to be $a=1.97$~fm,
and the time lattice spacing is $a_{t}%
=1.97$~fm$/c$.  The axial-vector coupling constant is $g_{A}=1.29$,
pion decay constant is$\ f_{\pi}=92.2$~MeV, pion mass is $m_{\pi}=m_{\pi^{0}}=134.98$~MeV,
and nucleon mass is $m=938.92$ MeV. We write
$\sigma_S$ with $S=1,2,3$ for the spin Pauli matrices, and $\tau_I$
with $I=1,2,3$ for the isospin Pauli
matrices.
We use dimensionless lattice units, where the physical quantities are multiplied
by powers of  the spatial lattice spacing $a$ to make dimensionless combinations.
We write $\alpha_t$ for the ratio $a_t/a$.  

The notation 
$\sum_{\langle{\bf n'\, n}\rangle}$
represents the summation over nearest-neighbor lattice sites of {\bf n}.
We use $\sum_{\langle{\bf n'\, n}\rangle_i}$ to indicate the sum over nearest-neighbor
lattice sites of {\bf n} along the $i^{\rm th}$ spatial axis.  Similarly,
 $\sum_{\langle\langle{\bf n'\, n}\rangle\rangle_i}$ is the sum
over   next-to-nearest-neighbor
lattice sites of {\bf n} along the $i^{\rm th}$ axis, and $\sum_{\langle\langle\langle{\bf
n'\, n}\rangle\rangle\rangle_i}$ is the sum
over   next-to-next-to-nearest-neighbor
lattice sites of {\bf n} along the $i^{\rm th}$ axis.
Our lattice system is defined on an $L \times L \times L$ periodic cube,
and so the
summations over ${\bf n'}$ are defined with periodic boundary conditions.

In our notation $a_{\rm NL}$ is a four-component spin-isospin column vector
while $a^\dagger_{\rm NL}$ is a four-component spin-isospin row vector.
For real parameter $s_{\rm NL}$,
we define the nonlocal annihilation and creation operators for each spin
and
isospin component of the nucleon,
\begin{align}
a_{\rm NL}({\bf n})&=a({\bf n})+s_{\rm NL}\sum_{\langle{\bf n'\, n}\rangle}a({\bf
n'}),\\
a^\dagger_{\rm NL}({\bf n})&=a^\dagger({\bf n})+s_{\rm NL}\sum_{\langle{\bf
n'\, n}\rangle}a^\dagger({\bf
n'}).
\end{align}
For spin indices $S=1,2,3,$ and isospin indices $I=1,2,3$, we define point-like
densities,
\begin{align}
\rho({\bf n})&= a^\dagger({\bf n}) a({\bf n}),
\\
\rho_{S}({\bf n})&=a^\dagger({\bf n})[\sigma_S] a({\bf
n}),
\\
\rho_{I}({\bf n})&=a^\dagger({\bf n})[\tau_I] a({\bf
n}),
\\
\rho_{S,I}({\bf n})&=a^\dagger({\bf n})[\sigma_S \otimes
\tau_I] a_{\rm }({\bf n}).
\end{align}
and also the
smeared nonlocal densities, 
\begin{align}
\rho_{\rm NL}({\bf n})&= a^\dagger_{\rm NL}({\bf n}) a_{\rm NL}({\bf n}),
\\
\rho_{S,\rm NL}({\bf n})&=a^\dagger_{\rm NL}({\bf n})[\sigma_S] a_{\rm NL}({\bf
n}),
\\
\rho_{I,\rm NL}({\bf n})&=a^\dagger_{\rm NL}({\bf n})[\tau_I] a_{\rm NL}({\bf
n}),
\\
\rho_{S,I,\rm NL}({\bf n})&=a^\dagger_{\rm NL}({\bf n})[\sigma_S \otimes
\tau_I] a_{\rm NL}({\bf n}).
\end{align}
For the leading-order short-range interactions we use
\begin{equation}
V_{\rm 0}=\frac{c_0}{2}\sum_{{\bf n'},{\bf n},{\bf n''}} : \rho_{\rm NL}({\bf
n'})
f_{s_{\rm L}}({\bf n'} - {\bf n})f_{s_{\rm L}}({\bf n} - {\bf n''}) \rho_{\rm
NL}({\bf
n''}):
\end{equation}
where $f_{s_{\rm L}}$ is defined for real parameter $s_L$ as
\begin{align}
f_{s_{\rm L}}({\bf n})& = 1 \; {\rm for} \; |{\bf n}| = 0, \nonumber \\
& = s_L \; {\rm for} \; |{\bf n}| = 1, \nonumber \\
& = 0 \; {\rm otherwise}.
\end{align}
The :: symbol indicates normal ordering, where the annihilation operators
are on the right-hand side and the creation operators are on the left-hand
side.   

The one-pion exchange interaction is given by
\begin{align}
V_{\rm OPE}=-\frac{g_A^2}{8f^2_{\pi}}\sum_{{\bf n',n},S',S,I}
:\rho_{S',I\rm }({\bf n'})f_{S'S}({\bf n'}-{\bf n})\rho_{S,I}({\bf n}):,
\end{align}
where $f_{S'S}$ is defined as
\begin{align}
f_{S'S}({\bf n'}{\bf -n})=\frac{1}{L^3}\sum_{\bf q}\frac{\exp[-i{\bf q}\cdot({\bf
n'}-{\bf n})-b_{\pi}{\bf q}^2]q_{S'}q_{S}}{{\bf q}^2 + m_{\pi}^2},
\end{align}
and each $q_S$ is an integer multiplied by $2\pi/L$.
 The parameter $b_{\pi}$ removes short-distance lattice artifacts
in the one-pion exchange interaction, and in this work we use the value $b_{\pi}=0.700$.
  We take the free lattice Hamiltonian to have the form \cite{Epelbaum:2010xt}
\begin{align}
H_{\rm free}= &\frac{49}{12m}\sum_{\bf n} a^\dagger({\bf n}) a({\bf n})-\frac{3}{4m}\sum_{{\bf
n},i}
\sum_{\langle{\bf n'}\,{\bf n}\rangle_i} a^\dagger({\bf n'}) a({\bf n}) \nonumber
\\
&+\frac{3}{40m}\sum_{{\bf
n},i}\sum_{\langle\langle{\bf n'}\,{\bf n}\rangle\rangle_i} a^\dagger({\bf
n'})
a({\bf n})-\frac{1}{180m}\sum_{{\bf
n},i}
\sum_{\langle\langle\langle{\bf n'}\,{\bf n}\rangle\rangle\rangle_i} a^\dagger({\bf
n'}) a({\bf n}).
\end{align}
The full leading-order (LO) lattice Hamiltonian can be written as 
\begin{align}
H_{\rm B} = H_{\rm free} + V_0 + V_{\rm OPE},
\end{align}
with $s_{\rm NL}=0.0800$, $s_{\rm L}=0.0800$, and $c_0=-0.1850$.   In tuning
our interactions here, we fit the parameters $s_{\rm NL}$, $s_{\rm L}$, and
$c_0$ to the average inverse scattering length and effective range of the
two $s$-wave channels, as well as the finite-volume energies of $^8$Be. 
The finite-volume energies for $^8$Be give a measure of the alpha-alpha scattering
length, which was emphasized in Ref.~\cite{Elhatisari:2016owd} as a sensitive
indicator correlated with the binding energies of medium-mass nuclei.

\subsection*{Nucleon-nucleon scattering}

\begin{figure}[!ht]
\centering
\caption{We plot LO lattice
phase shifts for proton-neutron scattering versus the center-of-mass relative
momentum.  For comparison
we also plot the phase shifts extracted from the Nijmegen partial wave analysis
\cite{Stoks:1993tb}.  In the first row, the data in panels {\bf a}, {\bf
b}, {\bf c}, {\bf d} correspond to ${^1s_0},{^3s_1}, {^1p_1}, {^3p_0}$ respectively.
In the second row, panels {\bf e}, {\bf f}, {\bf g}, {\bf h} correspond to
${^3p_1},{^3p_2},
{^1d_2}, {^3d_1}$ respectively.   In the third row, panels {\bf i}, {\bf
j}, {\bf k}, {\bf l} correspond to ${^3d_2},{^3d_3},
{\varepsilon_1}, {\varepsilon_2}$ respectively. \bigskip}
\includegraphics[scale=0.5]{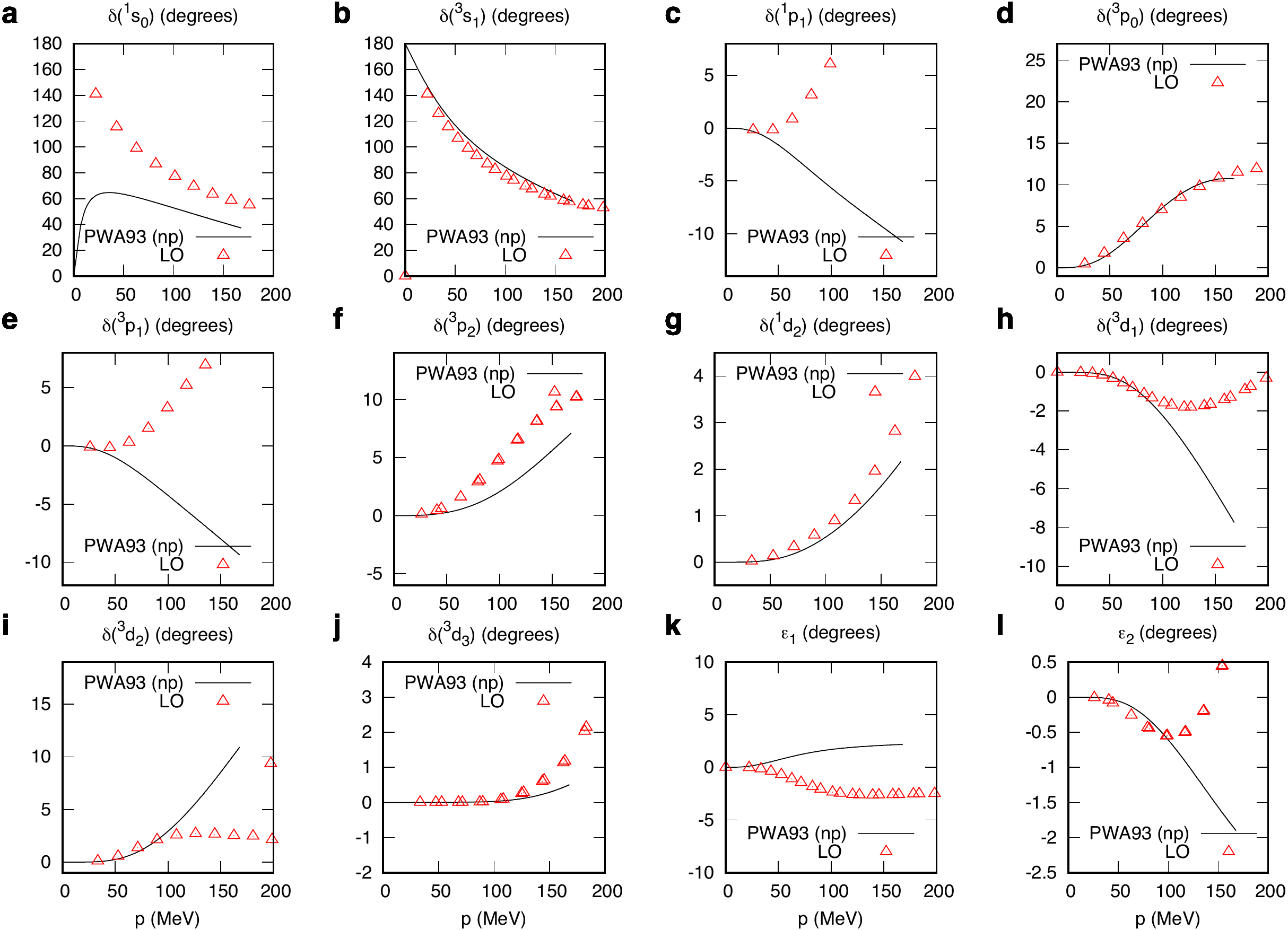}
\label{NN}
\end{figure}

The details of the nucleon-nucleon scattering calculations are given in 
Ref.~\cite{Elhatisari:2016owd}.  In Fig.~\ref{NN} we show the LO lattice
phase shifts for proton-neutron scattering versus the center-of-mass relative
momentum.
 For comparison we also present phase shifts from the Nijmegen
partial wave analysis \cite{Stoks:1993tb}.  In the first row, the
data in panels {\bf a}, {\bf b}, {\bf c}, {\bf d} correspond to ${^1s_0},{^3s_1},
{^1p_1}, {^3p_0}$ respectively. In the second row, panels {\bf e}, {\bf f},
{\bf g}, {\bf h} correspond to ${^3p_1},{^3p_2},
{^1d_2}, {^3d_1}$ respectively.   In the third row, panels {\bf i}, {\bf
j}, {\bf k}, {\bf l} correspond to ${^3d_2},{^3d_3},
{\varepsilon_1}, {\varepsilon_2}$ respectively.  As can been seen from  Fig.~\ref{NN},
the $^1s_0$ phase shift requires significant higher-order corrections. These
leading-order results are just the first step in the chiral
effective field theory expansion, and the phase shifts would be systematically
improved at each higher order, NLO, NNLO, etc.
While the behavior of the ${^1s_0}$ phase shift near threshold seems rather
poor, it requires only a rather small higher-order correction to reproduce
the proper ${^1s_0}$ phase shift.  We have checked this explicitly and it
is also one of the central themes in a recent paper on nuclear physics expanded
around the unitarity limit \cite{Konig:2016utl}.  The key point is that the
${^1s_0}$ phase shift depends strongly on small changes in the ${^1s_0}$
coupling strength because it sits very close to the unitarity limit where
the scattering length diverges.

\subsection*{Euclidean time projection and auxiliary-field Monte Carlo} 
The Euclidean time transfer matrix
$M$ is defined as the normal-ordered exponential of the lattice Hamiltonian
$H$ over one time lattice step,
\begin{align}
M = :\exp[-H\alpha_t]:. \label{Mtransfer}
\end{align}
We use an initial state $|\Psi_i\rangle $ and final state $|\Psi_f\rangle$
 that have nonzero overlap with the ground state nucleus of interest.  By
multiplying by
powers of $M$ upon $|\Psi_i\rangle$, we can project out the ground state.
 We compute projection
amplitudes of the form
\begin{equation}
Z_{f,i}(L_t) = \langle \Psi_f| M^{L_t}  |\Psi_i\rangle. 
\end{equation}
By calculating the ratio $Z_{f,i}(L_t)/Z_{f,i}(L_t-1)$ for large $L_t$ we
can
determine the ground state energy.  

It is useful however to first prepare
the initial state using a simpler transfer matrix $M_*$ that is a good approximation
to $M$.  We choose $M_*$ to be invariant under Wigner's
SU$(4)$ symmetry \cite{Wigner:1937}.  The SU(4) symmetry eliminates
sign oscillations  from auxiliary-field Monte Carlo simulations of $M_*$
\cite{Chen:2004rq,Lee:2007eu}.    
$M_*$ has the same form as $M,$ but the operator coefficients that violate
SU(4) symmetry are turned off.  We use $M_*$ as an approximate low-energy
filter by multiplying the initial and final states by $M_*$ some fixed
number of times, $L_t'$,
  \begin{equation}
Z_{f,i}(L_t) = \langle \Psi _f| M_*^{L'_t}   M^{L_t}   M_*^{L'_t}  |\Psi_i\rangle.
\label{Mstar_transfer}
\end{equation} 

We use auxiliary fields to generate the lattice interactions. \ The auxiliary
field method can be viewed as a Gaussian
integral formula which relates the exponential
of the two-particle density, $\rho^2$, to the integral of the exponential
of the one-particle density, $\rho$,\begin{equation}
{:\exp\left(-\frac{c\alpha_t}{2}\rho^2\right):}=\sqrt{\frac{1}{2\pi}}\int^{\infty}_{-\infty}ds
\, {:\exp \left(-\frac{1}{2}s^2 + \sqrt{-c\alpha_t}s\rho \right):}\;. \label{HS}
\end{equation}
The normal ordering symbol :: ensures that the operator products of the creation
and annihilation operators behave as classical anticommuting
Grassmann variables~\cite{Lee:2008fa}. We use this integral identity to
introduce auxiliary fields at every lattice site \cite{Hubbard:1959ub,Stratonovich:1958,Koonin:1986}.
The pion fields are treated in a manner similar to the auxiliary
fields. 
 
We couple the auxiliary field $s$ at time step $n_t$ to $\rho_{\rm NL}$ through
a convolution with the smearing function $f_{s_{\rm
L}}$.  The linear term in
the auxiliary field is
\begin{equation}
V^{(n_t)}_s = \sqrt{-c_{\rm 0}} \sum_{{\bf n,n'}}\rho_{\rm NL}({\bf n})f_{s_{\rm
L}}({\bf n} - {\bf n'})s({\bf
n'},n_{t}),
\end{equation}
and the quadratic term in the auxiliary field is
\begin{equation}
V^{(n_t)}_{ss} = \frac{1}{2} \sum_{{\bf n}}s^2({\bf
n},n_t).  
\end{equation}
For the one-pion exchange interaction, the gradient of the pion
field $\pi_I$ is coupled to
the point-like density $\rho_{S,I}$, 
\begin{equation}
V_{\pi}^{(n_t)}=\frac{g_A}{2f_{\pi}} \sum_{{\bf n,n'},S,I}\rho_{S,I}({\bf
n'})f^{\pi}_{S}({\bf n'-n)}\pi_{I}({\bf n},n_{t}),
\end{equation}
\begin{equation}
V_{\pi\pi}^{(n_t)}=\frac{1}{2}{} \sum_{{\bf n,n'},I}\pi_I({\bf
n'},n_{t})f^{\pi\pi}_{}({\bf n'-n)}\pi_{I}({\bf n},n_{t}),
\end{equation}
where $f^{\pi}_{S}$ ($S=1,2,3$) and $f^{\pi \pi}$ are defined as
\begin{equation}
f^{\pi}_{S}({\bf n'}{\bf -n})=\frac{1}{L^3}\sum_{\bf q}\exp[-i{\bf
q}\cdot({\bf
n'}-{\bf n})]q_S,
\end{equation} 
\begin{equation}
f^{\pi\pi}({\bf n'}{\bf -n})=\frac{1}{L^3}\sum_{\bf q}\exp[-i{\bf q}\cdot({\bf
n'}-{\bf n})+b_{\pi}{\bf q}^2]({\bf q}^2 + m_{\pi}^2).
\end{equation}Then the transfer matrix at leading order can be
written as an path integral,
\begin{align}
M =\int Ds^{(n_t)}D\pi^{(n_t)} \,M^{(n_t)},
\end{align}
where $Ds^{(n_t)}$ is the path integral measure for $s$ at time step $n_t$,
$D\pi^{(n_t)}$ is
the path integral measure for $\pi_I$ ($I = 1,2,3$) at time step $n_t$, and
\begin{align}
M^{(n_t)}= \,:\exp\left(-H_{\rm free}\alpha_t-V^{(n_t)}_s\sqrt{\alpha_t}-V^{(n_t)}_{ss}-V_{\pi}^{(n_t)}\alpha_t
-V_{\pi\pi}^{(n_t)}\alpha_t\right):.
\end{align}

In the projection Monte Carlo calculations we use the same procedure for
the initial states as discussed in Ref.~\cite{Elhatisari:2016owd}. Four nucleons
are inserted at each time step. For neutron-rich nuclei we also insert pairs
of spin-up and spin-down neutrons, and for proton-rich nuclei we insert pairs
of spin-up and spin-down protons.  For the calculations of ${^3}$H and ${^3}$He
we use an $L \simeq 16\,{\rm fm}$ periodic box, and for the rest of the nuclei
we use an $L \simeq 12\,{\rm
fm}$ periodic box.

\subsection*{Results for the ground state energies}

\begin{figure}[!ht]
\centering
\caption{We show the energy versus projection time for ${^3}$H and ${^3}$He.
Since the leading-order action is isospin invariant, the results are the
same
for the two nuclei.  The error bars indicate one standard deviation
errors from the stochastic noise of the Monte Carlo simulations, and
the line shows the extrapolation to infinite projection time. \bigskip}

\includegraphics[scale=0.4]{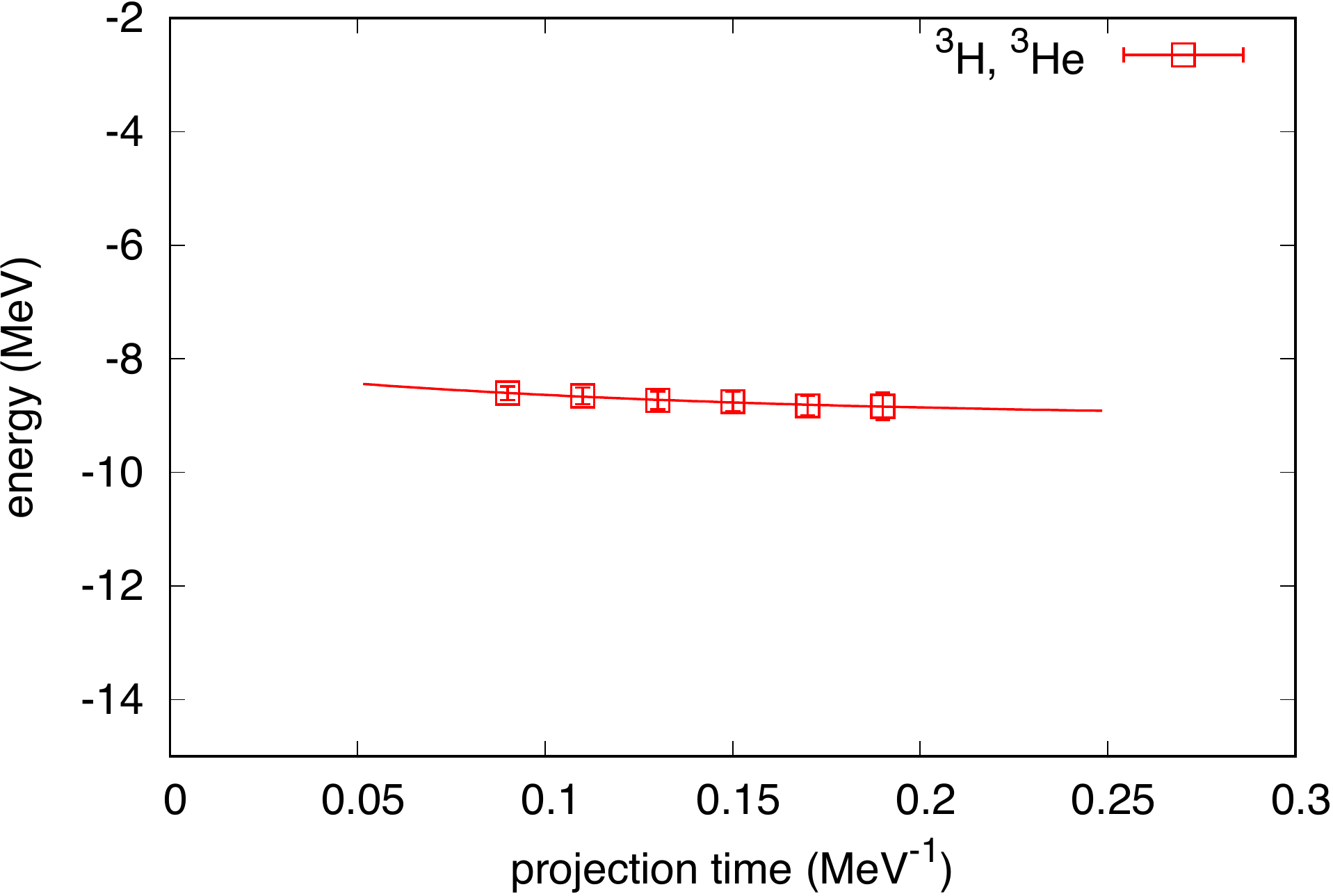}
\label{H3_new}
\end{figure}

\begin{figure}[!ht]
\centering
\caption{We show the energy versus projection time for the helium isotopes.
The error bars indicate one standard deviation
errors from the stochastic noise of the Monte Carlo simulations, and
the lines show extrapolations to infinite projection time. \bigskip}
\includegraphics[scale=0.4]{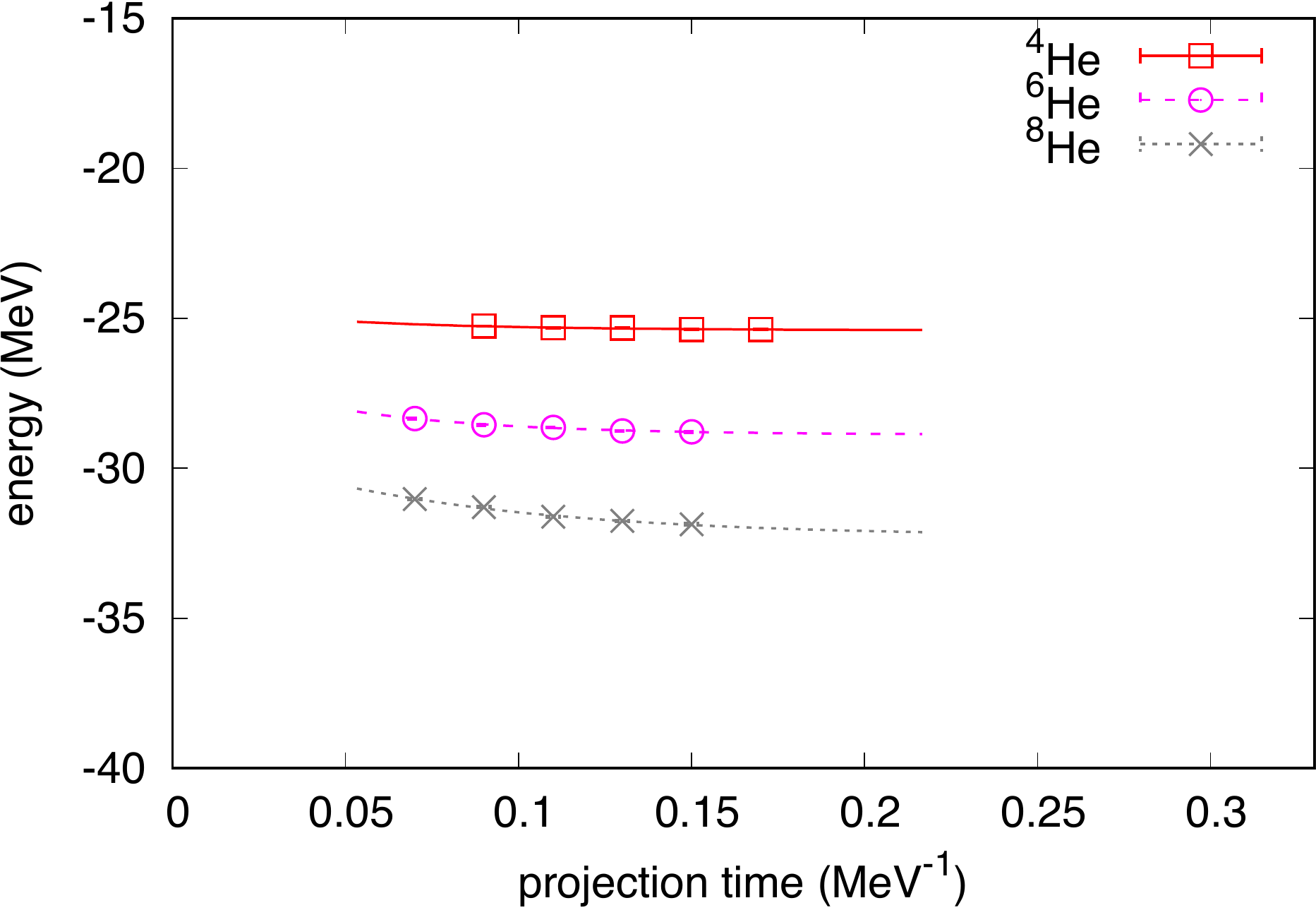}
\label{He_new}
\end{figure}

\begin{figure}[!ht]
\centering
\caption{We show the energy versus projection time for the beryllium isotopes.
The error bars indicate one standard deviation
errors from the stochastic noise of the Monte Carlo simulations, and
the lines show extrapolations to infinite projection time. \bigskip}
\includegraphics[scale=0.4]{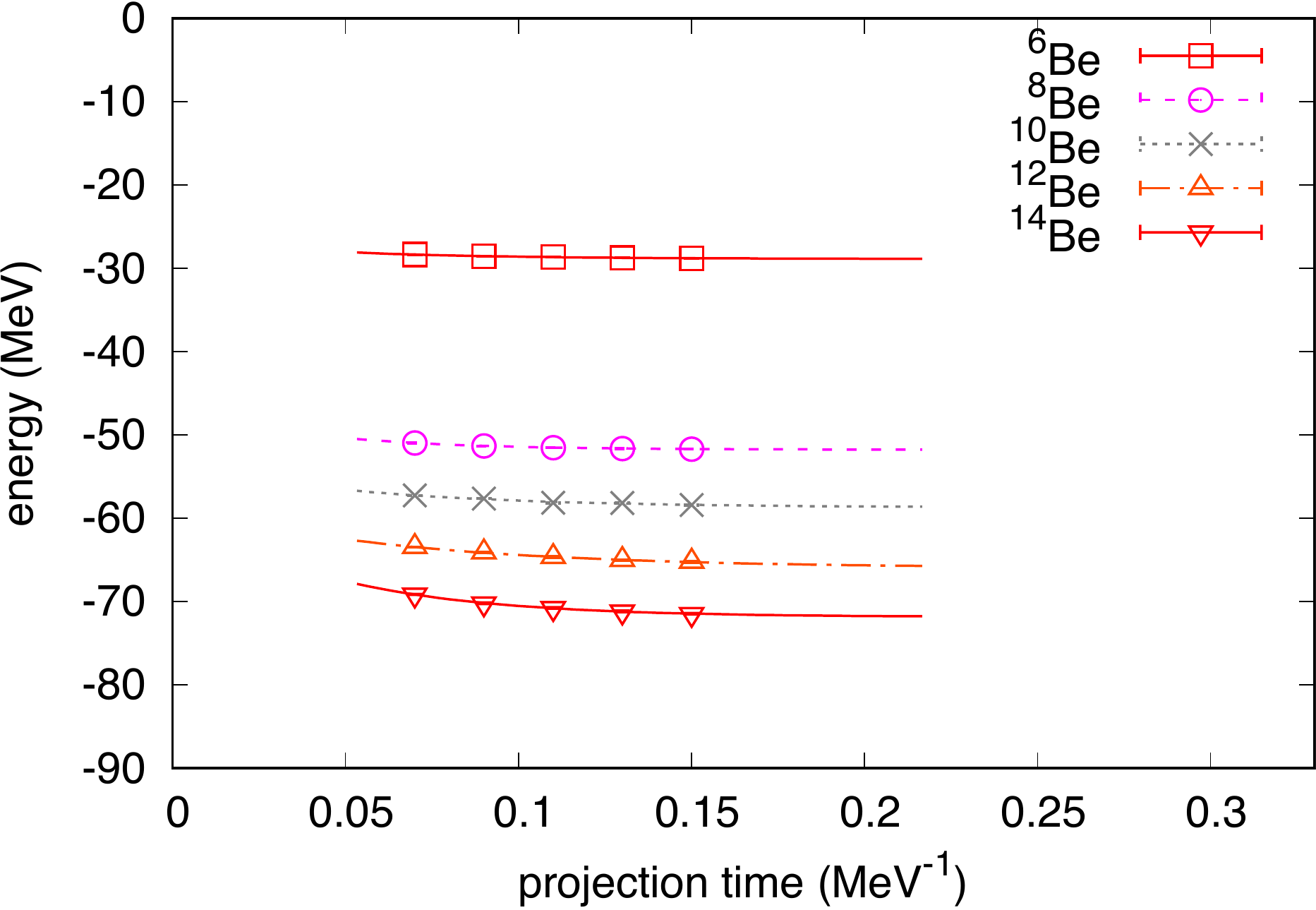}
\label{Be_new}
\end{figure}

\begin{figure}[!ht]
\centering
\caption{We show the energy versus projection time for the carbon isotopes.
The error bars indicate one standard deviation
errors from the stochastic noise of the Monte Carlo simulations, and
the lines show extrapolations to infinite projection time. \bigskip}
\includegraphics[scale=0.4]{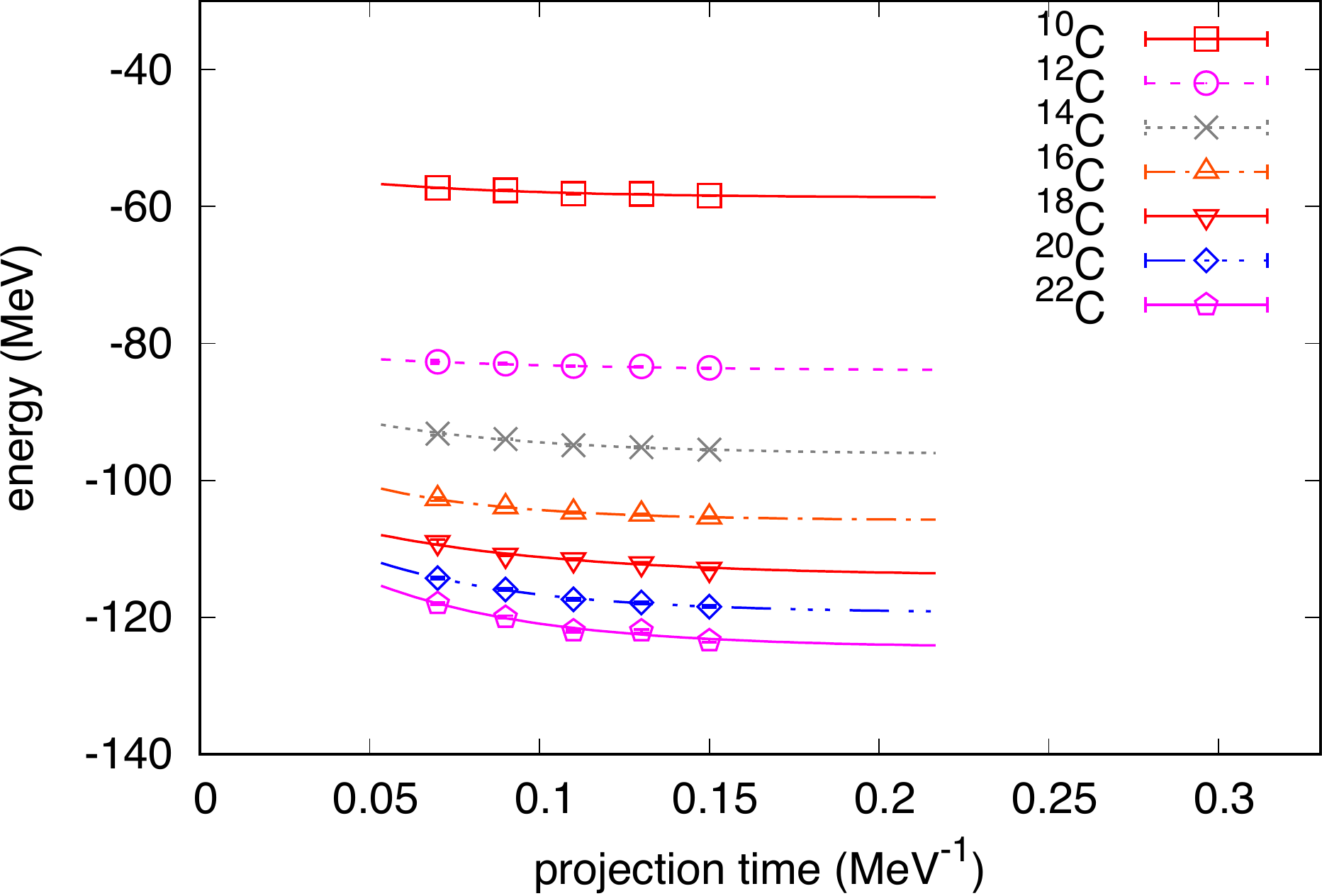}
\label{C_new}
\end{figure}

\begin{figure}[!ht]
\centering
\caption{We show the energy versus projection time for the oxygen isotopes.
The error bars indicate one standard deviation
errors from the stochastic noise of the Monte Carlo simulations, and
the lines show extrapolations to infinite projection time. \bigskip}
\includegraphics[scale=0.4]{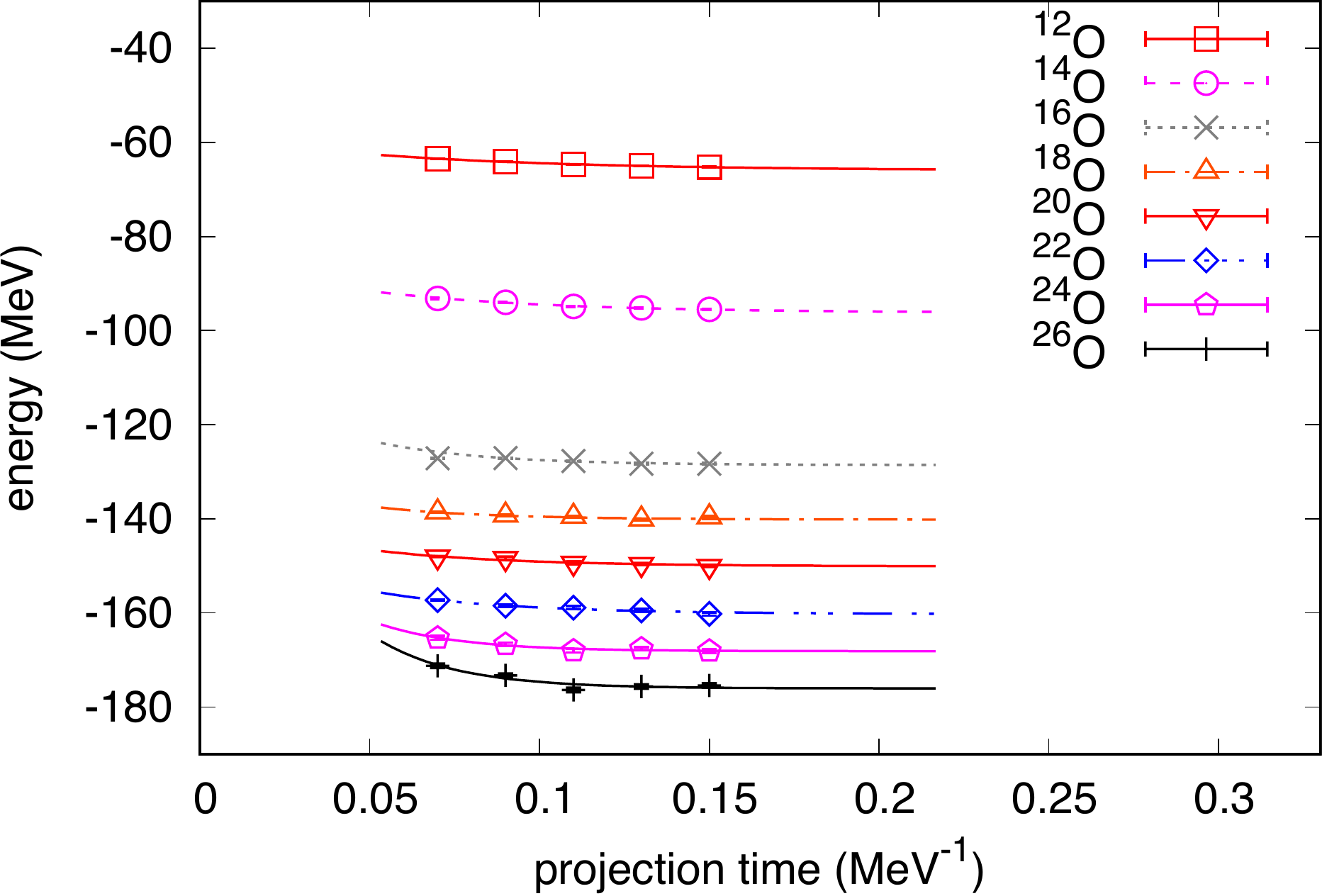}
\label{O_new}
\end{figure}

In Fig.~\ref{H3_new} we show the energy versus projection time for ${^3}$H
and ${^3}$He. The error bars indicate one standard deviation
errors due to the stochastic noise of the Monte Carlo simulations. The lines
are extrapolations to infinite projection time using the functional form
\begin{equation}
E(t) = E_{0} + c\exp[-\Delta E \, t],
\label{E_extrap}
\end{equation}
where $E_{0}$ is the ground state energy that we wish to determine.  The
results for the helium isotopes are shown in  Fig.~\ref{He_new}, the beryllium
isotopes in Fig.~\ref{Be_new}, the carbon isotopes in Fig.~\ref{C_new}, and
the oxygen isotopes in Fig.~\ref{O_new}.  

\subsection*{Results for $\rho_3$ and $\rho_4$}

We compute $\rho_3$ by inserting the operator
\begin{equation}
:\exp \left[\sum_{\bf n} \epsilon({\bf n})\rho({\bf n}) \right]:
\end{equation}
at the middle time step and taking three numerical derivatives with respect
to $\epsilon({\bf n})$ for
infinitesmally small $\epsilon({\bf n})$.  We then divide by $3!$ and sum
over {\bf n}.  For $\rho_4$ we compute four numerical derivatives with respect
to $\epsilon({\bf n})$, divide by $4!$, and sum over {\bf n}.

\begin{figure}[!ht]
\centering
\caption{We show $\rho_3$
versus projection time for the neutron-rich helium, beryllium, and carbon
isotopes.  The error bars indicate one
standard deviation errors from the stochastic noise of the Monte Carlo simulations,
and
the lines show extrapolations to infinite projection time. \bigskip } 
\includegraphics[scale=0.4]{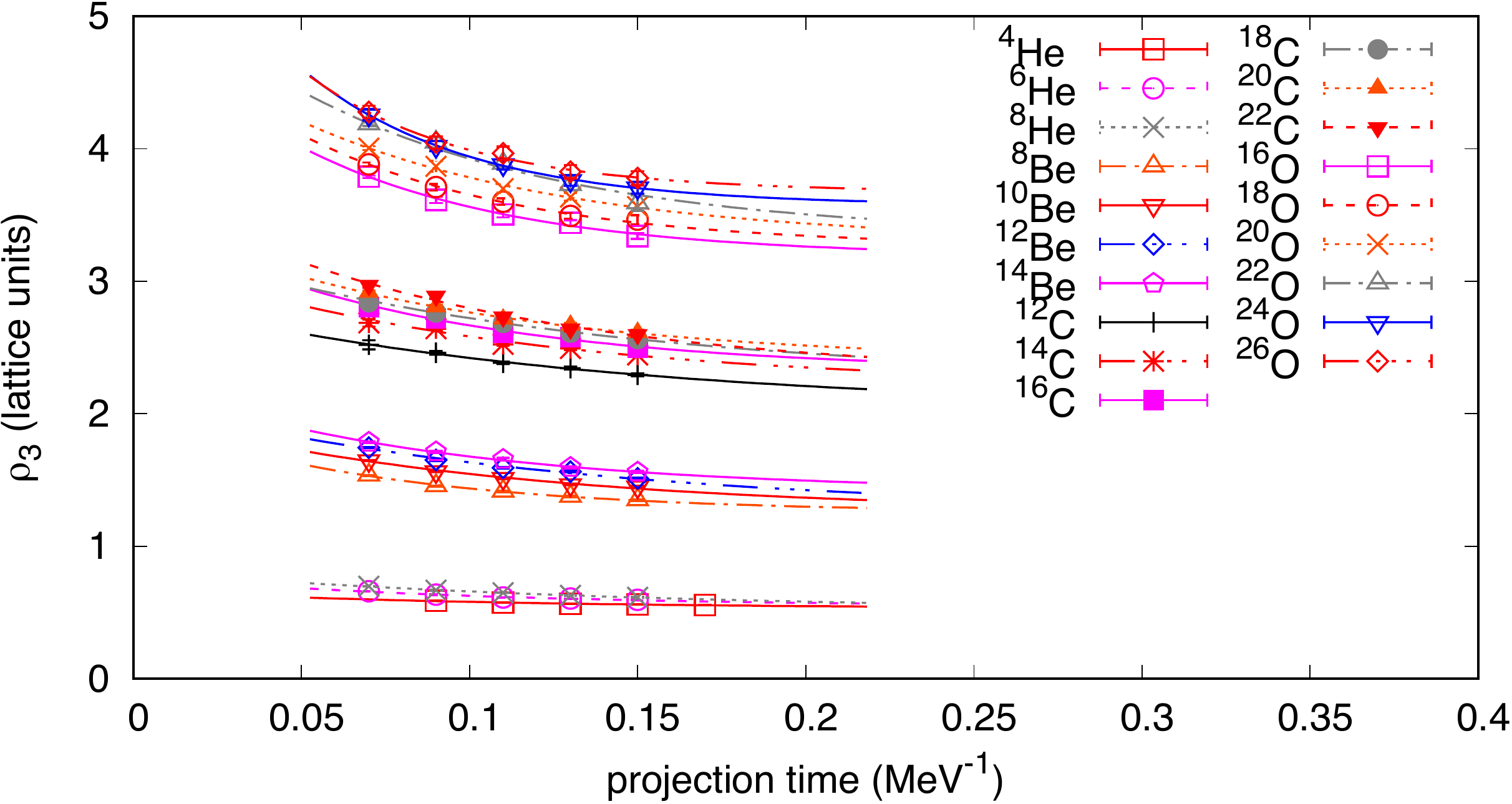}
\label{threeN_new}
\end{figure}

\begin{figure}[!ht]
\centering
\caption{We show $\rho_4$
versus projection time for the neutron-rich helium, beryllium, and carbon
isotopes.  The error bars indicate one
standard deviation errors from the stochastic noise of the Monte Carlo simulations,
and
the lines show extrapolations to infinite projection time. \bigskip } 
\includegraphics[scale=0.4]{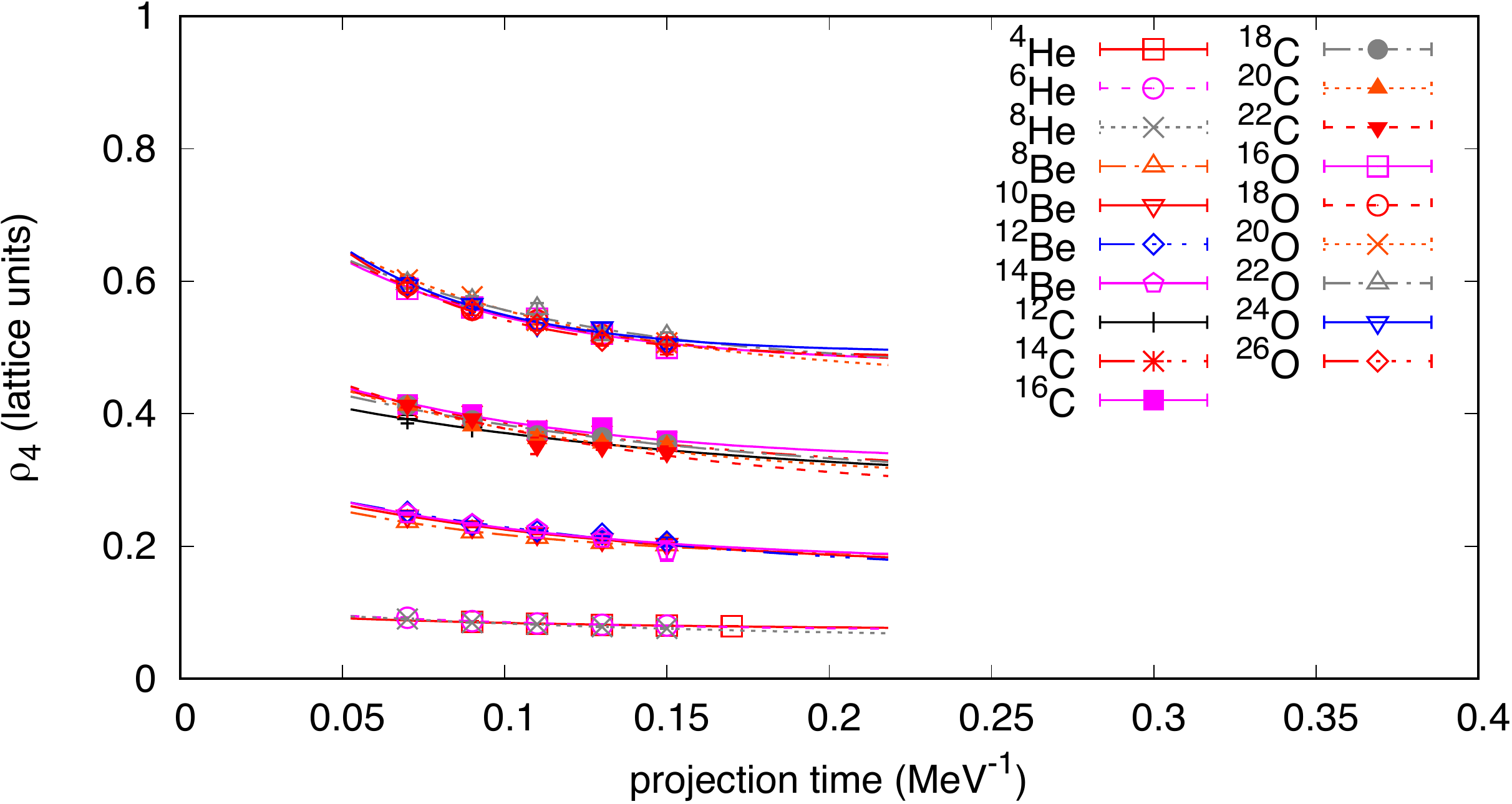}
\label{fourN_new}
\end{figure}

In Fig.~\ref{threeN_new} we show $\rho_3$ versus projection time for the
neutron-rich helium, beryllium, and carbon isotopes. The error bars indicate
one standard deviation
errors due to the stochastic noise of the Monte Carlo simulations. The lines
are extrapolations to infinite projection time using the functional forms\begin{align}
\rho_3(t) = \rho_3 + c_3\exp[-\Delta E \, t /2], \\
\rho_4(t) = \rho_4 + c_4\exp[-\Delta E \, t /2],
\label{rho_extrap}
\end{align}
where $\Delta E$ is determined from the ground state energy fit in 
Eq.~(\ref{E_extrap}).  The factor of $t/2$ rather than $t$ comes from the
fact that we are computing expectation values of $:\rho^3({\bf n})/3!:$ and
$:\rho^4({\bf n})/4!:$ inserted at the middle time step.  This leads to exponential
corrections from matrix elements connecting the ground state to the first
excited state, each of which are propagated for time duration $t/2$.  In
Fig.~\ref{fourN_new} we show $\rho_4$ versus projection time for the
neutron-rich helium, beryllium, and carbon isotopes.
 
\subsection*{Local cluster operators and operator product expansion}
Let us consider any short-distance three-nucleon operator of the form\begin{align}
X_3(\vec{r}) = 
\int d^3r_1 d^3r_2 d^3r_3 f(\vec{r}_1-\vec{r},\vec{r}_2-\vec{r},\vec{r}_3-\vec{r};\Delta
r):\rho(\vec{r}_1)\rho(\vec{r}_2)\rho(\vec{r}_3):, \label{short3}
\end{align}
where $f$ is a spatially-localized function with width  parameter $\Delta
r$.  
This operator can be expanded as a sum of local operator products \cite{Wilson:1969a,Zimmermann:1973a},
\begin{align}
X_3(\vec{r})=\sum_n O_n(\vec{r}) (\Delta r)^{d_n}c_n(\Delta r \Lambda), \label{OPE}
\end{align}
where $d_n$ is the momentum dimension of the operator $O_n(\vec{r})$, $\Lambda$
is the renormalization momentum scale, and $c_n$ gives the dependence on
$\Lambda$ through quantum loop effects.  The lowest possible value for $d_n$
is $9$ and is associated with the operator product $:\rho^3(\vec{r}):$. 
Other operators with the same quantum numbers as $X_3(\vec{r})$ have higher
dimension.  

Let us now consider the expectation value of $X_3(\vec{r})$ for the alpha
particle and for an arbitrary nucleus which we label $A$.  The ratio of these
expectation values is then
\begin{align}
\frac{\left< A \right| X_3(\vec{r}) \left| A \right>}{\left< \alpha \right|
X_3(\vec{r}) \left| \alpha \right>}=\frac{\left< A \right| :\rho^3(\vec{r}):
\left| A \right>}{\left< \alpha \right|
:\rho^3(\vec{r}): \left| \alpha \right>}+
\cdots=\frac{\rho_3}{\rho_{3,\alpha}}+ \cdots, \label{universal3}
\end{align} 
where the omitted terms are contributions from higher operators in Eq.~(\ref{OPE})
and therefore suppressed by powers of $\Delta r$. Similarly, we find that
for any short-distance four-nucleon operator  \begin{align}
X_4(\vec{r}) = 
\int d^3r_1 d^3r_2 d^3r_3 d^3r_4 &
f(\vec{r}_1-\vec{r},\vec{r}_2-\vec{r},\vec{r}_3-\vec{r},\vec{r}_4-\vec{r};\Delta
r):\rho(\vec{r}_1)\rho(\vec{r}_2)\rho(\vec{r}_3)\rho(\vec{r}_4):, \label{short4}
\end{align}
the ratio of expectation values is
\begin{align}
\frac{\left< A \right| X_4(\vec{r}) \left| A \right>}{\left< \alpha \right|
X_4(\vec{r}) \left| \alpha \right>}=\frac{\left< A \right| :\rho^4(\vec{r}):
\left| A \right>}{\left< \alpha \right|
:\rho^4(\vec{r}): \left| \alpha \right>}+
\cdots=\frac{\rho_4}{\rho_{4,\alpha}}+ \cdots. \label{universal4}
\end{align}

We can now turn Eq.~(\ref{universal3}) and Eq.~(\ref{universal4}) around
and conclude that the ratios ${\rho_3}/{\rho_{3,\alpha}}$ and ${\rho_4}/{\rho_{4,\alpha}}$
are independent of renormalization scale up to higher dimension corrections,
 
\begin{align}
& \frac{\rho_3}{\rho_{3,\alpha}}=\frac{\left< A \right| X_3(\vec{r}) \left|
A \right>}{\left< \alpha \right|
X_3(\vec{r}) \left| \alpha \right>} + \cdots, \\
& \frac{\rho_4}{\rho_{4,\alpha}}=\frac{\left< A \right| X_4(\vec{r}) \left|
A \right>}{\left< \alpha \right|
X_4(\vec{r}) \left| \alpha \right>} + \cdots.
\end{align}
It would be interesting to check this statement of model independence in
the future using a variety of different lattice and continuum {\it ab initio}
methods.

\subsection*{Local cluster operators as a measure of clustering}
Let us consider the short-distance operators $X_3(\vec{r})$ and $X_4(\vec{r})$
as defined in  Eq.~(\ref{short3}) and Eq.~(\ref{short4}) respectively.  As
the width of the spatial distributions $\Delta r$ becomes small, the  expectation
values of these short-distance operators will depend very strongly on the
amount of clustering present in the nucleus.  If the nucleus is a homogeneous
liquid of uncorrelated nucleons then 
\begin{eqnarray}
& \left< A \right| X_3(\vec{r}) \left| A \right> \sim \left( {\Delta
r}/{R_A} \right)^{6}, \\ & \left< A \right| X_4(\vec{r}) \left| A \right>
\sim \left( {\Delta
r}/{R_A} \right)^{9},
\end{eqnarray}
where $R_A$ is the radius of the nucleus $A$.  If on the other hand,
the nucleus is comprised of non-overlapping alpha clusters, then 
\begin{eqnarray}
& \left< A \right| X_3(\vec{r}) \left| A \right> \sim \left( {\Delta
r}/{R_\alpha} \right)^{6}, \\ & \left< A \right| X_4(\vec{r}) \left| A \right>
\sim \left( {\Delta
r}/{R_\alpha} \right)^{9}.
\end{eqnarray}
where $R_A$ is the radius of an alpha particle $A$.  Therefore if we measure
${\rho_3}/{\rho_{3,\alpha}}$ there is an enhancement by a factor of $(R_A/R_\alpha)^6$
if the nucleus is comprised of alpha clusters.  For ${\rho_4}/{\rho_{4,\alpha}}$
the enhancement factor is $(R_A/R_\alpha)^9$.

\subsection*{Pinhole algorithm}
Auxiliary-field Monte Carlo simulations are efficient for computing the quantum
properties of systems with attractive pairing interactions.  By the calculating
the exact quantum amplitude for each configuration of auxiliary fields, we
obtain the full set of correlations induced by the interactions.  However,
the exact quantum amplitude for each auxiliary field configuration involves
quantum states which are superpositions of many different center-of-mass
positions.  Therefore information about density correlations relative to
the center of mass is lost. The pinhole algorithm is a new computational
approach that allows for the calculation of arbitrary density correlations
with respect to the center of mass.  As this was not possible in all previous
auxiliary-field Monte Carlo simulations, adaptations of this technique should
have wide applications to hadronic, nuclear, condensed matter, and ultracold
atomic simulations. 

We let $\rho_{i,j}({\bf n})$ be the density operator for nucleons with spin
$i$ and isospin $j$ at lattice site {\bf n},
\begin{equation}
\rho_{i,j}({\bf n}) = a^\dagger_{i,j}({\bf n})a_{i,j}({\bf n}).
\end{equation}
We construct the normal-ordered $A$-body density operator
\begin{equation}
\rho_{i_1,j_1,\cdots i_A,j_A}({\bf n}_1,\cdots {\bf n}_A) = \; :\rho_{i_1,j_1}({\bf
n}_1)\cdots\rho_{i_A,j_A}({\bf
n}_A):.
\end{equation}
In the $A$-nucleon subspace, we note the completeness identity
\begin{equation}
\sum_{i_1,j_1,\cdots i_A,j_A}\sum_{{\bf n}_1,\cdots {\bf n}_A} 
\rho_{i_1,j_1,\cdots i_A,j_A}({\bf n}_1,\cdots {\bf n}_A) \; = A!.
\label{closure}
\end{equation}
Using the transfer matrices $M$ and $M_*$ defined in Eq.~(\ref{Mtransfer})
and Eq.~(\ref{Mstar_transfer}), in the pinhole algorithm we work with the
expectation value
\begin{equation}
 Z_{f,i}(i_1,j_1,\cdots i_A,j_A;{\bf n}_1,\cdots {\bf n}_A;L_{t})=\langle
\Psi _f| M_*^{L'_t}M^{L_t/2}      
\rho_{i_1,j_1,\cdots i_A,j_A}({\bf n}_1,\cdots {\bf n}_A)M^{L_t/2}M_*^{L'_t}
 |\Psi_i\rangle.
 \label{pinholeamp}
\end{equation}Due to the completeness identity Eq.~(\ref{closure}), the sum
of the expectation value in Eq.~(\ref{pinholeamp}) over ${\bf n}_1,\cdots
{\bf n}_A$ and $i_1,j_1,\cdots i_A,j_A$ gives $A!$ times the amplitude
\begin{equation}
 Z_{f,i}=\langle \Psi _f| M_*^{L'_t}M^{L_t}      
M_*^{L'_t}
 |\Psi_i\rangle. 
 \end{equation}

\begin{figure}[!ht]
\centering
\caption{A sketch of the pinhole locations and spin-isospin indices at time
$t = L_t a_t/2$.}
\includegraphics[scale=0.5]{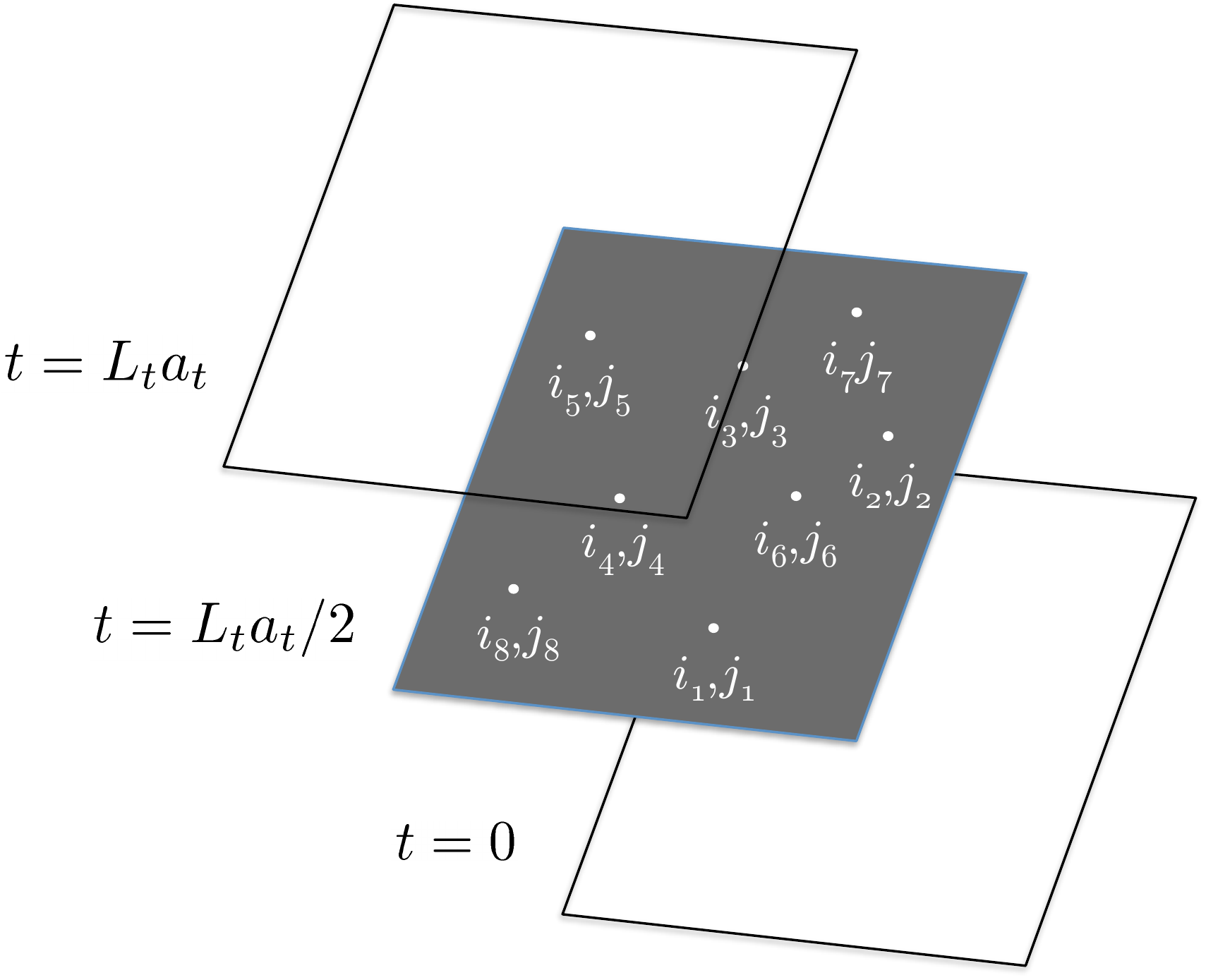}
\label{pinhole_sheet}
\end{figure}

The quantities $Z_{f,i}(i_1,j_1,\cdots i_A,j_A;{\bf n}_1,\cdots {\bf n}_A)$
and $Z_{f,i}$ are computed using Monte Carlo simulations with auxiliary fields.
 Within the auxiliary-field framework, the pinhole locations ${\bf n}_1,\cdots
{\bf n}_A$ and spin-isospin indices $i_1,j_1,\cdots i_A,j_A$ are sampled
by Metropolis updates \cite{Hastings:1970aa}, while the auxiliary fields
are sampled by the hybrid Monte Carlo algorithm \cite{Duane:1987de,Gottlieb:1987mq}.
In Fig.~\ref{pinhole_sheet} we show a sketch of the pinhole locations and
spin-isospin indices for the operator  $\rho_{i_1,j_1,\cdots i_A,j_A}({\bf
n}_1,\cdots {\bf n}_A)$ inserted at
time $t = L_t a_t/2$.  We obtain the ground state expectation value by extrapolating
to the limit of infinite projection time. We compute the path integrals
\begin{align}
Z_{f,i}(i_1,j_1,\cdots i_A,j_A;{\bf n}_1,\cdots {\bf n}_A;L_{t})=  \int
Ds^{}D\pi^{ }\langle
\Phi _f(s,\pi)|\rho_{i_1,j_1,\cdots i_A,j_A}({\bf n}_1,\cdots {\bf n}_A)^{}
 |\Phi _i(s,\pi)\rangle ,
 \label{pinholeint}
\end{align}
where $DsD\pi$ is the path integral measure for all time steps of the auxiliary
field $s$ and pion field $\pi$, and\begin{align}
|\Phi _i(s,\pi)\rangle & = M^{(L'_t+L_{t}/2-1)}\cdots
M^{(L'_{t})}M_*^{(L'_t-1)}\cdots M_*^{(0)}
 |\Psi _i \rangle, \nonumber \\
\langle\Phi _f(s,\pi) | &= \langle \Psi _f |M_*^{(2L'_t+L_t-1)}\cdots M_*^{(L'_t+L_{t})}M^{(L'_t+L_{t}-1)}\cdots
M^{(L'_{t}+L_{t}/2)}
 .
 \end{align}

We perform importance sampling of the path integral in Eq.~(\ref{pinholeint})
according to the absolute value of the integrand, 
\begin{equation}
A(s,\pi ;i_1,j_1,\cdots i_A,j_A;{\bf n}_1,\cdots {\bf n}_A;L_{t})=|\langle
\Phi _f(s,\pi)|\rho_{i_1,j_1,\cdots i_A,j_A}({\bf n}_1,\cdots {\bf n}_A)^{}
 |\Phi _i(s,\pi)\rangle|.
\end{equation} 
The complex phase of the integrand is treated as an observable that is accumulated
to give a total sum over all selected configurations. In the pinhole algorithm
we alternate the auxiliary
field and pion field updates with updates of the spin-isospin indices and
pinhole locations. For fixed indices $i_1,j_1,\cdots i_A,j_A$ and pinhole
locations ${\bf n}_1,\cdots {\bf n}_A$ we use the hybrid Monte Carlo algorithm
\cite{Gottlieb:1987mq,Duane:1987de} to update the auxiliary field and pion
field.  This is the same method used in previous nuclear lattice simulations,
and the details of the implementation can be found in Ref.~\cite{Lee:2008fa,Lee:2016fhn}.

The spin-isospin indices $i_1,j_1,\cdots i_A,j_A$ and are updated using the
Metropolis algorithm \cite{Metropolis:1953am}.  We propose a new set of indices
$i'_1,j'_1,\cdots i'_A,j'_A$ by randomly reassigning  the spin and isospin
for some of the nucleons. We select a random number $r$ uniformly distributed
between $0$ and $1$ and accept the new indices if\begin{equation}
r < \left| \frac{A(s,\pi ;i'_1,j'_1,\cdots i'_A,j'_A;{\bf n}_1,\cdots {\bf
n}_A;L_{t})}
{A(s,\pi ;i_1,j_1,\cdots i_A,j_A;{\bf n}_1,\cdots {\bf n}_A;L_{t})}
 \right|.
\end{equation}
We also choose new pinhole locations ${\bf n}'_1,\cdots
{\bf n}'_A$ by randomly displacing one of the pinhole locations by one lattice
unit. We select another random number $r$ uniformly distributed
between $0$ and $1$ and accept the new pinhole locations if\begin{equation}
r < \left| \frac{A(s,\pi ;i_1,j_1,\cdots i_A,j_A;{\bf n}'_1,\cdots {\bf
n}'_A;L_{t})}
{A(s,\pi ;i_1,j_1,\cdots i_A,j_A;{\bf n}_1,\cdots {\bf n}_A;L_{t})}
 \right|.
\end{equation}
In this manner we update the auxiliary and pion fields, spin-isospin indices,
and pinhole locations.     

\subsection*{Density correlations}

For spatial lattice spacing $a$, the coordinates ${\bf r}_i$ of each nucleon
on the lattice is an integer vector ${\bf n}_i$ times $a$.  We do not consider
mass differences between protons and neutrons in these calculations.  Since
the center of mass is a mass-weighted average of $A$ nucleons with the same
mass, the center-of-mass position ${\bf r}_{\rm CM}$ is an integer vector
${\bf n}_{\rm CM}$ times $a/A$. Therefore the density distribution has a
resolution scale that is $A$ times smaller than the lattice spacing.  In
order to determine the center-of-mass position ${\bf r}_{\rm CM}$, we minimize
the squared radius
\begin{equation}
\sum_i \left|{\bf r}_{\rm CM} - {\bf r}_i \right|^2,
\end{equation} 
where each term $\left|{\bf r}_{\rm CM} - {\bf r}_i \right| $ is minimized
with respect to all periodic copies of the separation distance on the lattice.
We comment that the tails of the proton and neutron density distributions
are determined from the asymptotic properties of the $A$-body wave function,
which
have been derived in a recent paper \cite{Konig:2017krd} for interactions
with finite range.

\begin{figure}[!ht]
\centering
\caption{We sketch a typical configuration of the protons (red) and neutrons
(blue) in $^{12}$C, with the arrows indicating up and down spins.  The triangle
of spin-up protons is indicated by the orange triangle. \label{alpha_triangle}}
\includegraphics[scale=0.5]{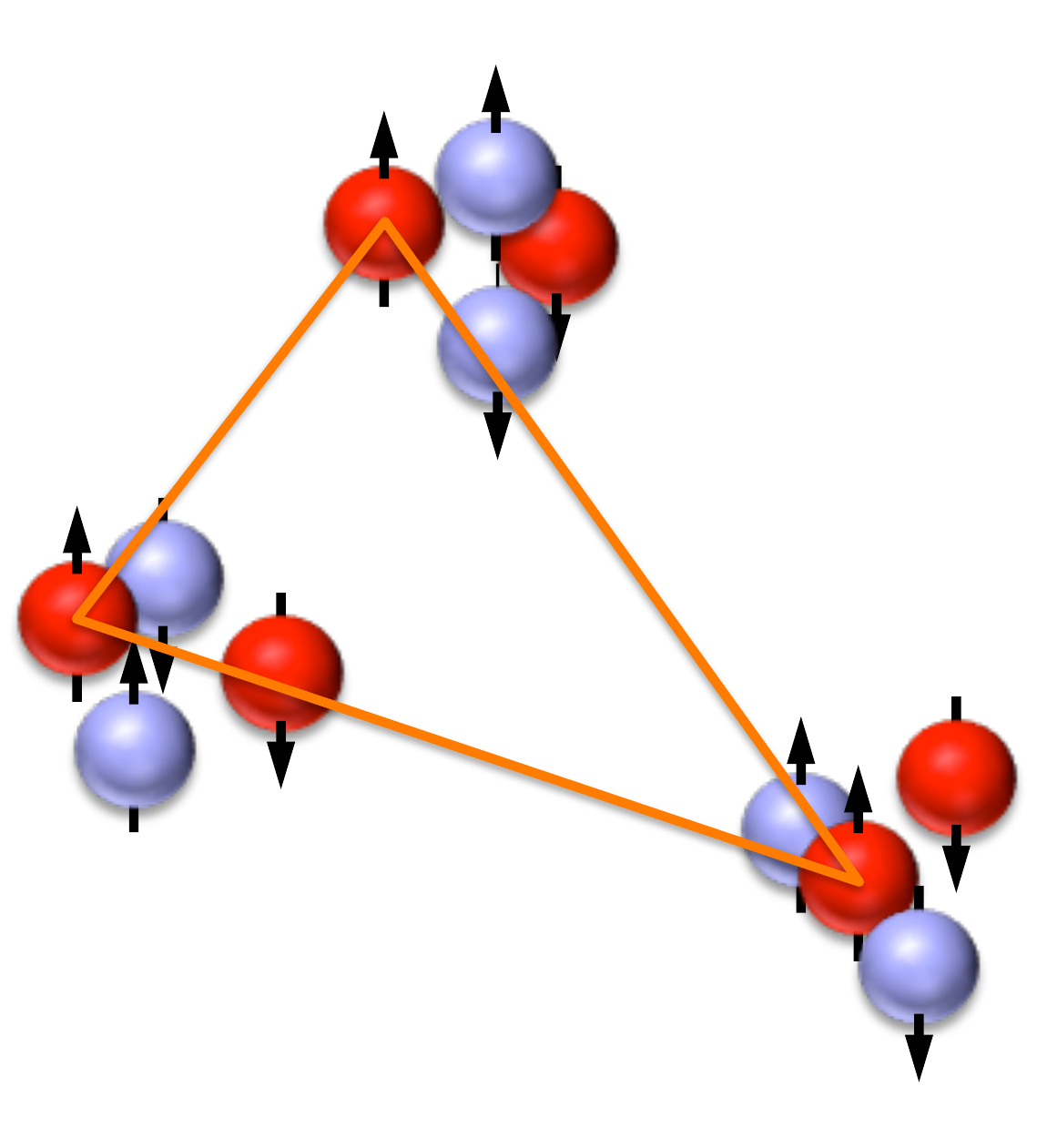}
\end{figure}

As discussed in the main text, from the $A$-body density information we can
view the triangular shapes formed by the three spin-up protons in the carbon
isotopes.  The positions of the three spin-up protons serve as a measure
of the alpha cluster geometry. In Fig.~\ref{alpha_triangle} we sketch a typical
configuration of the protons (red) and neutrons (blue) with the arrows indicating
up and down spins in $^{12}$C.  The three spin-up protons form the vertices
of a triangle, and this is indicated by the orange
triangle in Fig.~\ref{alpha_triangle}.  When collecting the lattice simulation
data, we rotate the triangle so that the longest side
lies on the $x$-axis.  We also rescale the triangle so the longest side has
length one, and flip the triangle, if needed, so that the third spin-up
proton is in the upper half of the $xy$-plane. 

\subsection{Form factors and radii}

\begin{figure}[!ht]
{\centering
\caption{The magnitude of the $^{12}$C electric form factor, $|F(q)|$, versus
momentum transfer $q$ in units of fm$^{-1}$.  The error bars indicate one
standard deviation errors from the stochastic noise of the Monte Carlo simulations.
For comparison we also show experimental results \cite{Sick:1970ma}. \label{C12_form}}
\includegraphics[scale=0.5]{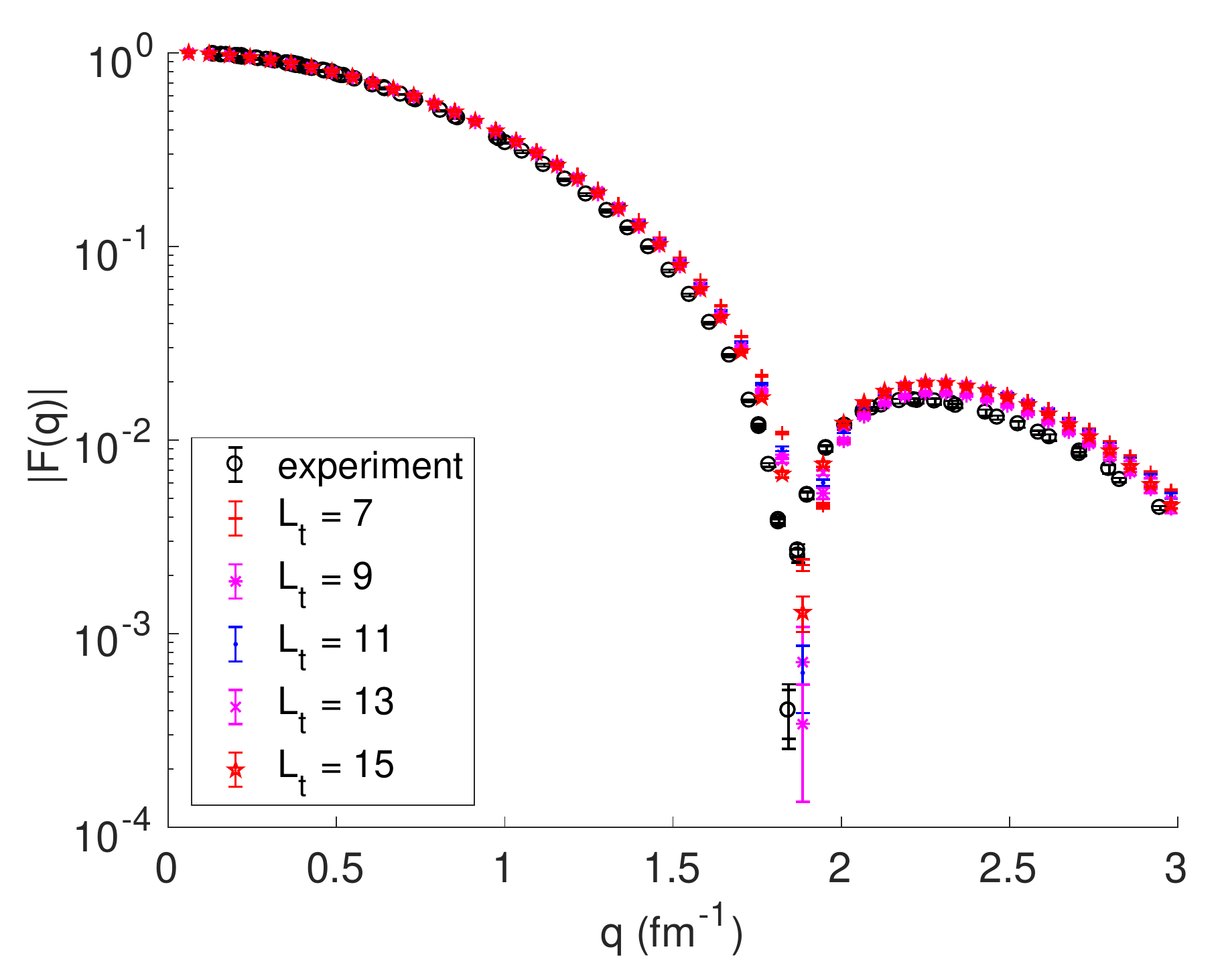}}
\end{figure}

From the density distribution of the protons relative to the center of mass,
we compute the Fourier transform to determine the electric form factor, $F(q)$,
where $q$ is the momentum transfer.  In order to reduce systematic errors
due to the lattice spacing, we perform a least squares fit of the density
distribution using a two-parameter Fermi model,
\begin{equation}
\rho(r) = \frac{\rho_0}{1+e^{(r-c)/z}},
\end{equation}
and then Fourier transform to momentum space.
The results are shown in Fig.~\ref{C12_form}. 

\begin{figure}[!ht]
\centering
\caption{The magnitude of the $^{14}$C electric form factor, $|F(q)|$, versus
momentum transfer $q$ in units of fm$^{-1}$.  The error bars indicate one
standard deviation errors from the stochastic noise of the Monte Carlo simulations.
For comparison we also show experimental results \cite{Kline:1973pi}.}
\label{C14_form}
\includegraphics[scale=0.5]{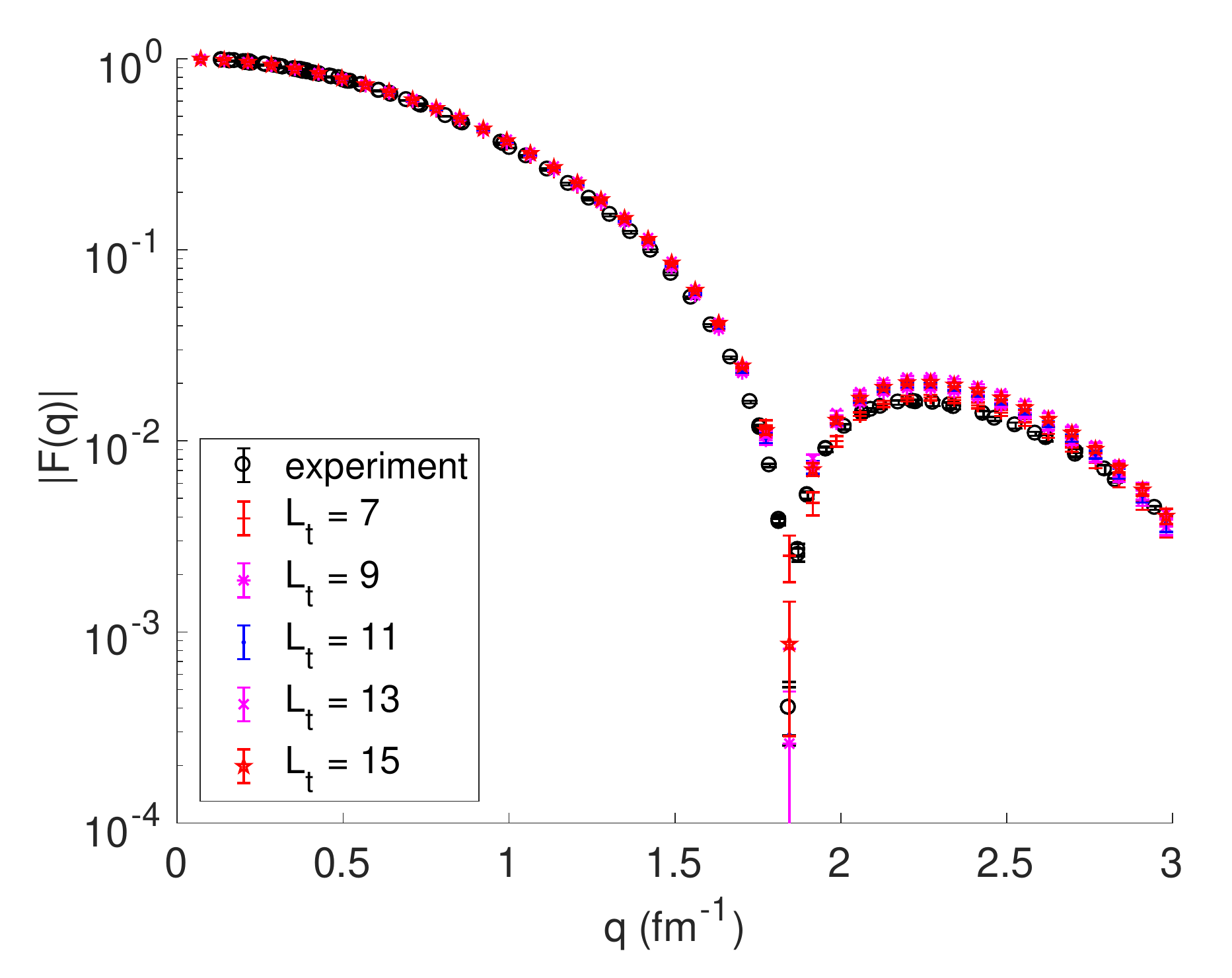}
\end{figure}

From the density distribution of protons and neutrons, we also compute the
root-mean-square (rms) radius for the proton and neutron distributions at
leading order.  The results are shown in Table~\ref{radii}. The shown error
bars include Monte Carlo errors as well as errors due to extrapolation to
infinite projection time. For comparison we show the rms charge radius observed
in electron scattering experiments. We find reasonable agreement between
the $^{12}$C and $^{14}$C proton radii at leading order and the corresponding
observed charge radii.   

\centering
\begin{table}[!ht]
\caption{Observed charge radii from electron scattering and proton and neutron
radii at leading order. \label{radii}}
\begin{tabular}{|c|c|c|c|}\hline
nucleus & observed charge radius & proton radius (LO) & neutron radius
(LO)  \\\hline
$^{12}$C   & 2.472(16) fm \cite{Schaller:1982}, 2.481(6) fm \cite{Sick:1982wq}
  & 2.40(7) fm & 2.39(5)  fm  \\\hline
$^{14}$C & 2.497(17) fm \cite{Schaller:1982} & 2.43(7) fm & 2.56(7) fm  \\\hline
$^{16}$C & --- & 2.46(10) fm  & 2.65(8) fm  \\\hline
\end{tabular}
\end{table}

\bibliographystyle{apsrev}
\bibliography{References}

\end{document}